\documentclass[12pt,preprint]{aastex}

\usepackage{graphicx}

\newcommand{\simgt}{\lower.5ex\hbox{$\;\buildrel>\over\sim\;$}}
\newcommand{\simlt}{\lower.5ex\hbox{$\;\buildrel<\over\sim\;$}}

\newcommand{\hst}{{\it HST}}
\newcommand{\ppc}{{\it pc/pixel}}
\newcommand{\hii}{H{\sc II}}

\slugcomment{Submitted to ApJ}

\shorttitle{Black Hole Fueling}
\shortauthors{Hunt \& Malkan }

\begin{document}

\title{Circumnuclear Structure and Black Hole Fueling:
\hst/NICMOS Imaging of 250 Active and Normal Galaxies}

\author{L. K. Hunt}
\affil{Istituto di Radioastronomia-Sezione Firenze, 
Largo E. Fermi 5, I-50125 Firenze, Italy}
\email{hunt@arcetri.astro.it}

\and

\author{M. A. Malkan}
\affil{Department of Astronomy, University of California, Los Angeles, CA, USA}
\email{malkan@astro.ucla.edu}

\begin{abstract}
Why are the nuclei of some galaxies more active than others?  If most galaxies
harbor a central massive black hole, the main difference is probably in how well
it is fueled by its surroundings.  We investigate the hypothesis that such a
difference can be seen in the detailed circumnuclear morphologies of 
galaxies using several quantitatively defined features,
including bars, isophotal twists, boxy and disky isophotes, and strong
non-axisymmetric features in unsharp masked images.
These diagnostics are applied to 250 high-resolution
images of galaxy centers obtained in the near-infrared with NICMOS on \hst.
To guard against the influence of possible biases and selection effects, we
have carefully matched samples of Seyfert 1, Seyfert 2, LINER, starburst and normal 
galaxies in their basic properties, taking particular care to
ensure that each was observed with a similar average scale
($10-15$ parsecs per pixel).  
Several morphological differences among our five different
spectroscopic classifications emerge from the analysis.
The \hii/starburst galaxies show the strongest deviations from smooth
elliptical isophotes, while the normal galaxies and LINERS have the
least disturbed morphology. The Seyfert 2 galaxies have significantly more
twisted isophotes than any other category, and the early-type Seyfert 2s
are significantly more disturbed than the early-type Seyfert 1s.
The morphological differences between Seyfert 1s and 2s suggest that more
is at work than simply the viewing angle of the central engine. They may
correspond to different evolutionary stages.
\end{abstract}


\section{Introduction}

The established theoretical explanation for Seyfert activity
requires that all active galactic nuclei (AGNs) must have formed
central massive black holes (BHs) and are now fueling them, presumably with gas
from the host galaxy that has lost most of its orbital angular momentum.
However, BHs are not exclusive to AGNs, since quiescent ones
are now found in most, if not all, galaxies with massive spheroids. 
The BH mass seems to depend on certain properties of the bulge in which
it resides \citep{fm00,gebhardt00,tremaine02,marconi}.
But only a fraction of galaxies \citep{hoagn,sdss} host AGN, and it 
is not understood why some galaxies contain AGNs, while the majority
do not.

The prevailing explanation implicates the efficiency of nuclear fueling. 
Although the gas available for fueling the AGN may play an important
role, there is as yet only weak evidence for 
larger gas fractions in Seyfert galaxies (e.g., \citealt{huntetal99}).
Most research has instead concentrated on the mechanism by which disk 
gas loses its angular momentum, thereby becoming available to feed the BH.
The main focus has been on galactic bars, which are efficient at transporting 
gas on kpc scales, but are unable to funnel gas inward to strictly nuclear 
scales (pc--tens of pcs); to overcome this,
nested bars were proposed as a possible mechanism \citep{shlosman89}.
Nevertheless, Seyfert galaxies have not shown an excess of of large-scale bars 
\citep{moles,mcleod,ho97,mulchaey97,regan99,huntmalkan99,regan99,marquez00},
although Seyferts may host bars on smaller spatial scales \citep{knapen00,laine02}.
It has also been suggested that only type 2 Seyferts show an excess
of bars \citep{maiolino97}. 

Another focus has been on tidal interactions, since these are also
theoretically viable mechanisms for inward gas transport \citep{hernquist,barnes}.
However, Seyferts are found neither preferentially
in interacting systems nor with an excess of companions \citep{fuentes,derobertis},
although this too is still a point of debate \citep{dultzin99,krongold}. 

Hence, it appears that large-scale galactic structure in Seyfert host galaxies
(with the possible exception of their disks, see \citealt{huntetal99}), 
has little bearing on the creation and fueling of the AGN.  
It is possible however that the causes (or effects) of the
AGN can be found on small spatial scales, such as those available to 
\hst. 
Indeed, using a sample of \hst/NICMOS images,
\citet{martini99} and \citet{regan99} suggested that
nuclear spirals are responsible for fueling AGNs, but later work
did not confirm this \citep{martini03b}. 
With much of the same imaging data, \citet{laine02} found
a significant excess of bars in Seyferts on all spatial scales,
including circumnuclear bars as revealed by \hst.
Again there is no consensus about AGN fueling mechanisms even with 
careful analyses of virtually the same high-resolution images.

In this paper, we readdress the issue of AGN fueling on small 
spatial scales with the largest sample of active and non-active 
galaxies ever compiled in this context. 
We start with virtually all galaxies imaged with \hst/NICMOS in the 
{\it F160W} ({\it H} band, 1.6 $\mu$m) filter, but carefully construct
subsamples separated according to activity type so as to eliminate potential biases.  
Our study differs from previous ones in several important ways:
{\it (i)}~to maximize sample size,
we incorporate data from all three NICMOS cameras, taking care to ensure
that the images have similar spatial scales; 
{\it (ii)}~LINERs and \hii/starburst galaxies are included in 
the analysis in order
to investigate possible evolutionary trends and the importance of 
star formation; 
{\it (iii)}~active and non-active samples are constructed
to have comparable {\it medians and ranges} in parsec-to-pixel spatial scale, 
{\it B}-band luminosity, distance, inclination, and Hubble type;
{\it (iv)}~objective techniques are used to detect isophotal twists, bars,
boxy/disky isophotes, banana or heart-shaped isophotes, and non-axisymmetric
structure in unsharp masked images.
Except for bars and visually identified non-axisymmetric structure, 
none of these features has been analyzed previously.

Morphology studies such as this one are best conducted in
the near-infrared (NIR) bands in general, and the {\it F160W} filter 
in particular. 
The NIR effectively traces the bulk of the stellar mass because of 
its sensitivity to the cooler stars which dominate evolved stellar populations. 
This means that the massive stars associated with recent star formation
are less apt to disturb morphology, allowing us to use the NIR images
as a rough proxy for mass distributions.
NIR wavelengths are also much less
affected by dust extinction than optical bands, and suffer very little gas
contamination.  

Our initial sample of 250 galaxies includes all nuclear activity classes, from 
``normal'' (non-active)
and \hii/starburst, to LINER, and Seyfert galaxies of both types, so that we can 
perform a comparative analysis, and investigate how nuclear morphology of the 
galaxy
influences, if at all, the creation and fueling of an AGN.

\section{The Images \label{sec:images}}

We have acquired \hst/NICMOS F160W images of 30 Seyfert galaxies in a GO
Snapshot Program (ID 5479, Malkan PI). 
Our GO snapshots have been augmented
with other F160W snapshot images from the \hst\ archive acquired with
all three NICMOS cameras (Pogge-ID 7867, Mulchaey
-ID 7330, Sparks-ID 7919, Stiavelli-ID 7331, Peletier-ID 7450).  

\subsection{Target Selection \label{sec:targets}}

Our Snapshot targets were a subset of Seyfert galaxies with z$\leq$0.015 listed in
the 1993 Veron-Cetty and Veron AGN catalog \citep{veron},
imaged in a WFPC2 Snapshot program (ID 5479, Malkan PI).
Many were selected because of their unusual properties.  
For example, several
galaxies classified as Seyfert 1s had no detectable point source in our WFPC2 images,
while 
several galaxies classified as Seyfert 2s showed strong point sources. 
About half of the targets --particularly those with strong point-like nuclei-- were observed
with the highest resolution camera, NIC-1 (0.043 arcsec pixels; 11 arcsec field-of-view: FOV).
The remainder were observed with the medium resolution camera, NIC-2 (0.075 
arcsec pixels; 19.2 arcsec FOV).

The other observed sets of Seyfert galaxies were based on similar criteria.
Using NIC-1,
Pogge imaged 23 of the CfA Seyfert 2s \citep{huchraburg} known from WFPC2 imaging to
have dusty centers. 
With NIC-2,
Mulchaey observed 104 Seyfert and comparison normal galaxies, selected 
from the Revised Shapley-Ames (RSA) catalog,
excluding those with $v\,>\,5000$\,km\,s$^{-1}$ and axial ratios $>\,0.35$.

Normal spiral samples were more diversified, but comprise mainly early Hubble types.
A large sample of nearby, mostly normal galaxies was imaged in {\it F160W} snapshots
with the NIC-3 camera by Sparks, randomly selected from the RSA according to
\hst\ scheduling convenience.
An atlas of these images
has been published by \citet{boker}.  
Normal Sa to Sbc galaxies were observed by Stiavelli with NIC-2.
This sample was selected from the UGC \citep{nilson} and the ESOLV \citep{lauberts}
catalogs,
and excluded galaxies with $v\,>\,2500$\,km\,s$^{-1}$, inclinations $>\,75^\circ$, 
and systems with known bars.  
Additional spirals imaged with NIC-2 by Peletier
were selected from a $B$-magnitude limited sample \citep{balcellspeletier}, 
with inclinations $>\,50^\circ$, and of early Hubble type (S0-Sbc).

\subsection{Image Processing \label{sec:reduction}}

We re-reduced all images using the {\it STSDAS/calnica}\footnote{STSDAS is the
Space Telescope Science Data Analysis System.} routine 
using the ``best available"
calibration frames for bias subtraction, dark subtraction, flatfielding and bad pixel
identification, rather than the frames that were originally used.
The images were then corrected, quadrant-by-quadrant, for the unpredictable drifts
in the bias level which produces the well-known ``pedestal"--a positive or negative
ghost of the flatfield which remains in the reduced image.  The four bias level
corrections (one for each quadrant) that must be made are determined by
an iterative process in the pedestal removal algorithm
of van der Marel (see \url{http://www.stsci.edu/$_{\tilde{\,}}$marel/software/pedestal.html}).  

Excepting program 7919, the observations were made with two, three, or 
four equal exposures shifted in a small
dither pattern (an ``L" shape for the triple exposures, and a square for the
quadruple exposures).  
We determined the exact shifts with the IRAF\footnote{IRAF is the Image Analysis and 
Reduction Facility made available to the astronomical community by the National Optical
Astronomy Observatory, which is operated by AURA, Inc., under
contract with the U.S. National Science Foundation.} task {\it xregister},
and then summed the dithered images using the {\it imshift} and {\it combine} tasks 
in IRAF.
Bad pixel masks were generated from the Data Quality flags, and
augmented by hand when necessary after visual inspection of the
final combined image.
Representative galaxy images are shown in Figures \ref{fig:imgnrm}, \ref{fig:imghii},
\ref{fig:imglin}, \ref{fig:imgsy2}, and \ref{fig:imgsy1} (see $\S$\ref{sec:samples}).

\subsection{Photometric Calibration \label{sec:phot}}

A key advantage of infrared imaging over almost all optical imaging obtained
with \hst\ is the large dynamic range over which flux measurements remain linear.
Particularly for Seyfert 1s with bright nuclei, 
even in relatively short exposures most WFPC2 images suffer
saturation which cannot be corrected.
Photometry from the 500-second exposures in \citet{mgt} (hereafter MGT)
is suspect for point sources brighter than {\it V}=19 
(see \url{http://www.astro.ucla.edu/$_{\tilde{\,}}$malkan/mgt.txt}).

Fortunately, all of our NICMOS images have linear flux scales even into the
centers of bright Sy1 nuclei.  This allows us to obtain accurate
photometry, perform image deconvolution, and model fitting to the
central brightness distributions, none of which are practical for
most WFPC2 images.  We used the {\it F160W} zeropoints of 21.667, 21.826, and 
21.566 respectively, for the NICMOS Cameras 1, 2 and 3, which puts our magnitudes
on the H (Vega) scale 
(see \url{http://www.stsci.edu/hst/nicmos/performance/photometry/keywords.html}).

\begin{figure*}[t]
\centering
{\includegraphics[width=0.9\linewidth, bb=30 145 595 730]{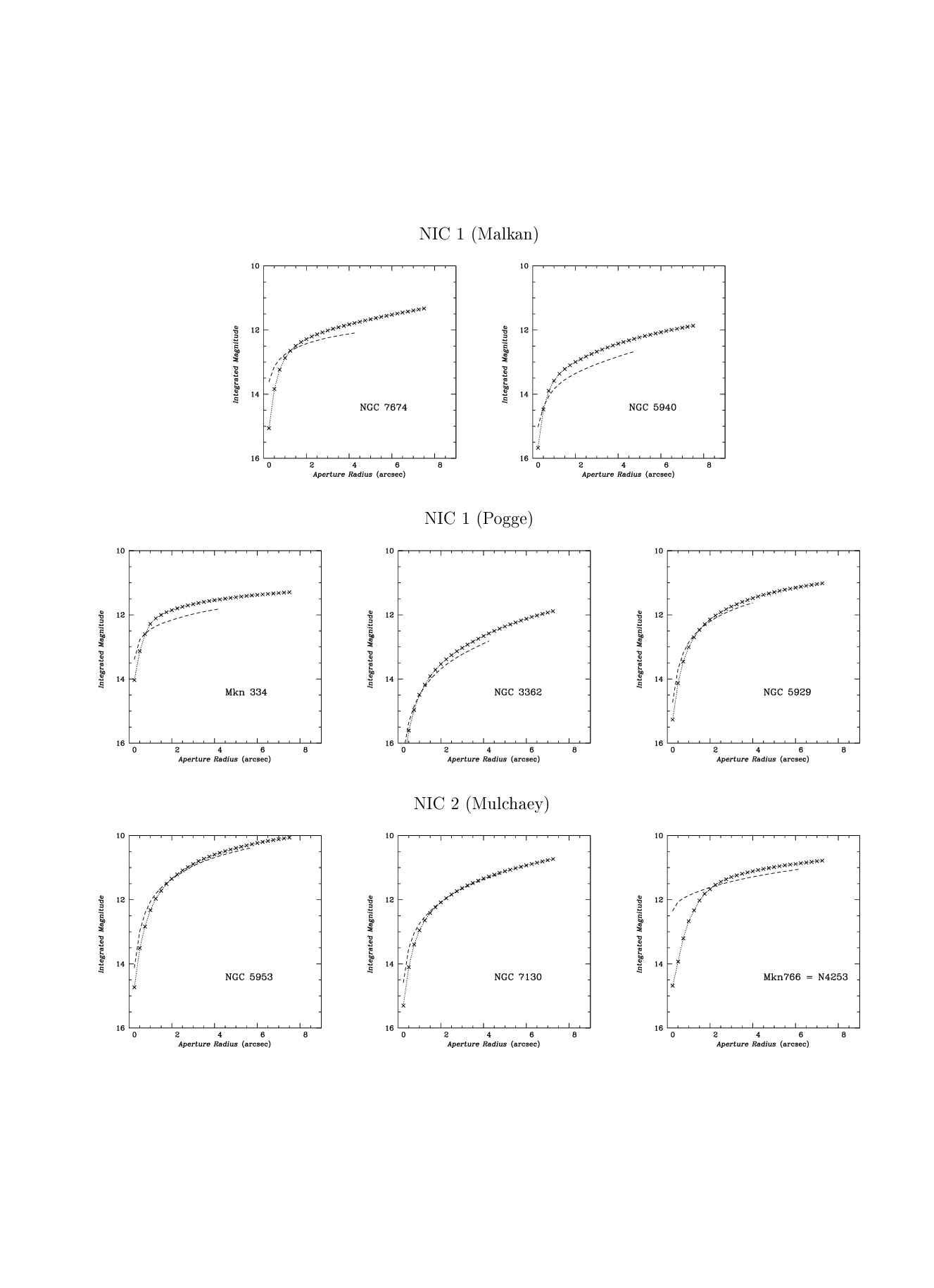}}
\caption{Curves of growth for ground-based images in the $H$-band
filter and \hst/NICMOS with {\it F160W}. Dashed lines show
NICMOS photometry, and $\times$ mark the ground-based.
\label{fig:phot}}
\end{figure*}

We made a detailed comparison shown in Figure \ref{fig:phot}
of the {\it F160W} photometric scale using our own 
ground-based {\it H}-band images of 8 of the sample galaxies \citep{12umatlas}.
The growth curves of magnitude versus aperture diameter agree to within 5\%
based on the transformations for NIC-2.
This good agreement was obtained assuming a zero sky level, hence this assumption
was maintained for all cameras.  In general, the comparison of our NICMOS photometry
to our ground-based measurements shows little evidence for any time-variability
of the nucleus at 1.6$\mu$m, roughly consistent with the findings of \citet{quillen}.
The NIC-1 H magnitudes are about $\sim$0.1 fainter at all radii, but we have
not corrected the data for this systematic offset, since
we are interested in morphology rather than absolute calibration. 

\placefigure{fig:phot}

\section{The Matched Samples \label{sec:samples}}

Given the variety of selection preferences that led to the observation of these
250 objects, they cannot be considered fair samples of local active or
normal galaxies.
We therefore have carefully selected the five activity-type subsamples so as to
mitigate potential biases.

The galaxies were classified  
according to their optical spectra, following NED, as
normal (non-active galaxies),
\hii-region/starbursts (\hii), LINERs, Seyfert 2s (Sy2s), or Seyfert 1s (Sy1s).
The physical foundation of some of these categories is not absolutely clear.
For example, galaxies having ``low ionization" line emission (``LINERs")
may be a heterogeneous class which includes some galaxies with recent
star formation and possible associated wind outflows with shocked gas,
as well as some genuine low-power active galaxies, with relatively
weak central nonstellar engines.
If a galaxy was classified as having both an AGN and \hii-region-like spectra,
we placed it in the more ``active'' category (LINER or Seyfert).

A small fraction of the images from the comprehensive target set was discarded a priori
because of bad pointing, which shifted the center of the galaxy partly or 
entirely off the detector.  
About two dozen galaxy images were rejected from further 
analysis because they are too irregular or do not contain any clear nucleus. 

\subsection{Parameter Control and Sample Construction}

The next step in the sample construction process was to constrain several
physical parameters.
Galaxies are complex objects which span very wide ranges in virtually every 
observable, and our aim is the mitigation of 
selection biases which could distort the results 
by producing spurious ``differences" among samples.
Nevertheless, there are intrinsic differences between galaxies which
host Seyfert nuclei and their normal non-active counterparts.
Seyfert galaxies tend to be of early Hubble type \citep{moles}, 
and more luminous \citep{huchraburg}.
Optically selected Seyferts also tend to avoid edge-on systems
\citep{keel}, although infrared-selected Seyferts are less affected
by this bias \citep{huntmalkan99}.
Our approach here is to maximize the size 
of the sample to optimize statistical significance, while at the same time, 
minimize the sample differences which could result in biases.
The distributions of a given parameter in each activity sample
were examined, and the extrema eliminated.
Sample medians and ranges were then recomputed, and this process was repeated
until extrema and medians are similar for all activity classes. 

This approach provides several advantages over previous studies.
First, instead of creating paired samples (e.g., \citealt{martini03b}),
we retain as many galaxies as possible since we require {\it statistical}
similarity of the samples rather than individual matching among
galaxies. 
Second, rather than modifying binning intervals to match
distributions (e.g., \citealt{laine02}),
we constrain the ranges and medians of the samples, so their statistical
properties should be more robust.
Third, the resulting Seyfert samples maintain the principal characteristics
of Seyfert galaxies, namely slightly higher luminosity and earlier Hubble
type than normal galaxies.
We however ensure that these differences are as small as possible, and
analyze subsamples where necessary to verify that they are 
{\it not} the cause of any differences (e.g., $\S$\ref{sec:twists}, \ref{sec:3theta}, 
\ref{sec:usmresiduals}). 

Our highest priority is to study samples of galaxies with various categories
of nuclear activity with {\it equivalent physical spatial resolution}, as
measured by parsecs per pixel.
This consideration is especially important given our use of all three
NICMOS cameras which differ by more than a factor of four in pixel size.
Because of the \hst\ diffraction limit at 1.6$\,\mu$m of 0.17\arcsec,
it is also necessary to constrain distance.
Blue luminosity constraints were applied in order to eliminate
possible Malmquist biases and trends of structure with luminosity,
independently of activity type.
Hubble type is also checked so as to ensure that potential differences
are not simply a function of galaxy morphology.
Finally, we checked large-scale galaxy inclination, 
so as to exclude highly-inclined systems in which circumnuclear morphology 
may not be easy to measure.
Bar class as given in RC3 was also checked, but not constrained.

We started with 250 galaxies, and ended by eliminating 85 of them, so that
the final sample consists of 165 galaxies, 47 of which are non-active;
this set of samples will be designated hereafter as MS (Matched Samples).
From this, we also constructed two additional sets.
In the first, distance is
further constrained to be $\leq$\,80\,Mpc (denoted as DMS, Distance-Matched Samples),
and in the second, we require inclination $i$ to be $<$ 70$^\circ$ (denoted
as IMS, Inclination-Matched Samples). 
Details of the sample matching are given in the Appendix, 
together with a list of the 85 galaxies eliminated.  
The medians and ranges of the parameters for each of the 
matched sample sets is reported in Table \ref{tab:medians}.
For each activity class, the first line reports medians and standard
deviations, and the second line the range.
Col. 8. is an exception to this where only percentages of bar class are given.
The values in parentheses are quartiles, not standard deviations.
The final samples of galaxies separated by activity type
are listed in Table \ref{tab:sample}, with NED designations, redshifts,
RC3 classifications, and optical major and minor axes and magnitudes.

\clearpage
\begin{deluxetable}{lccccccc} 
\tablecolumns{8} 
\tableheadfrac{0.05} 
\tabletypesize{\scriptsize}
\tablewidth{0pt}
\tablenum{1}
\tablecaption{Matched Sample Properties\label{tab:medians}}
\tablehead{ 
\multicolumn{1}{c}{Activity} & 
\multicolumn{1}{c}{Number\tablenotemark{a} } & 
\multicolumn{1}{c}{Distance} &
\multicolumn{1}{c}{Resolution} & 
\multicolumn{1}{c}{Absolute} &
\multicolumn{1}{c}{RC3 Type} & 
\multicolumn{1}{c}{$\cos(i)$} &
\multicolumn{1}{c}{\% SB}\\ 
\multicolumn{1}{c}{Class} & &
\multicolumn{1}{c}{[Mpc]} &
\multicolumn{1}{c}{[Pc/pixel]} & 
\multicolumn{1}{c}{Magnitude} & & &
\multicolumn{1}{c}{\% SAB}\\ 
\colhead{(1)} &  \colhead{(2)} & 
\colhead{(3)} & \colhead{(4)} &
\colhead{(5)} & \colhead{(6)} & \colhead{(7)} & \colhead{(8)} }
\startdata
Normal         & 47           & 27.8 (4.8)  & 10.4 (2.6) & $-19.3$ (1.0)  & 3.0 (3.0) & 0.45 (0.25) & 32\%\\
               &              & 11.4$-$67.4 & 5.2$-$24.5 & $-18.5- -21.4$ & $-2 - 9$  & 0.13$-$0.90 & 32\%\\
               & 47$^*$       & 27.8 (4.8)  & 10.4 (2.6) & $-19.3$ (1.0)  & 3.0 (3.0) & 0.45 (0.25) \\
               &              & 11.4$-$67.4 & 5.2$-$24.5 & $-18.5- -21.4$ & $-2 - 9$  & 0.13$-$0.90 \\
               & 32$^\dagger$ & 25.0 (5.2)  & 10.1 (1.8) & $-19.2$ (0.6)  & 3.0 (1.5) & 0.59 (0.18) \\
               &              & 11.4$-$38.6 & 5.2$-$15.6 & $-18.5- -21.4$ & $-2 - 9$  & 0.35$-$0.90 \\
\\
\hii/starburst & 14           & 27.0 (12.2) & 12.6 (3.2) & $-19.7$ (0.5)  & 4.0 (1.5) & 0.66 (0.19) & 42\%\\
               &              & 10.9$-$75.3 & 4.0$-$15.7 & $-18.5- -21.2$ & $0 - 7$   & 0.37$-$0.93 & 42\%\\
               & 14$^*$       & 27.0 (12.2) & 12.6 (3.2) & $-19.7$ (0.5)  & 4.0 (1.5) & 0.66 (0.19) \\
               &              & 10.9$-$75.3 & 4.0$-$15.7 & $-18.5- -21.2$ & $0 - 7$   & 0.37$-$0.93 \\
               & 14$^\dagger$ & 27.0 (12.2) & 12.6 (3.2) & $-19.7$ (0.5)  & 4.0 (1.5) & 0.66 (0.19) \\
               &              & 10.9$-$75.3 & 4.0$-$15.7 & $-18.5- -21.2$ & $0 - 7$   & 0.37$-$0.93 \\
\\
LINER          & 22           & 18.5 (5.0)  & 7.5 (2.7)  & $-19.8$ (0.7)  & 2.0 (1.5) & 0.64 (0.18) & 10\%\\
               &              & 10.6$-$55.1 & 3.9$-$20.0 & $-18.4- -22.0$ & $-4 - 5$  & 0.31$-$1.00 & 38\%\\
               & 22$^*$       & 18.5 (5.0)  & 7.5 (2.7)  & $-19.8$ (0.7)  & 2.0 (1.5) & 0.64 (0.18) \\
               &              & 10.6$-$55.1 & 3.9$-$20.0 & $-18.4- -22.0$ & $-4 - 5$  & 0.31$-$1.00 \\ 
               & 20$^\dagger$ & 18.5 (4.7)  & 7.5 (3.8)  & $-19.8$ (0.6)  & 2.0 (1.5) & 0.65 (0.18) \\ 
               &              & 10.6$-$55.1 & 3.9$-$20.0 & $-18.6- -22.0$ & $-4 - 5$  & 0.36$-$1.00 \\
\\
Sy 2           & 55           & 37.4 (17.9) & 12.9 (5.1) & $-20.4$ (0.7)  & 2.0 (2.0) & 0.73 (0.17) & 37\%\\
               &              & 11.8$-$117.4 & 4.9$-$25.9 & $-18.6- -21.7$ & $-4 - 5$ & 0.18$-$1.00 & 37\%\\
               & 51$^*$       & 36.0 (14.8) & 12.5 (4.5) & $-19.9$ (0.7)  & 2.0 (2.0) & 0.73 (0.16)  \\
               &              & 11.8$-$71.3 & 4.9$-$25.9 & $-18.6- -21.7$ & $-4 - 5$  & 0.18$-$1.00  \\
               & 51$^\dagger$ & 38.3 (19.9) & 13.2 (5.2) & $-20.4$ (0.7)  & 2.0 (3.0) & 0.74 (0.15) \\
               &              & 11.8$-$117.4 & 4.9$-$25.9 & $-18.6- -21.7$ & $-4 - 5$ & 0.37$-$1.00 \\
\\
Sy 1           & 27           & 48.5 (38.2) & 15.2 (6.1) & $-20.3$ (0.5)  & 3.0 (3.0) & 0.67 (0.22) & 32\%\\
               &              & 11.4$-$119.4 & 3.3$-$24.9 & $-19.0- -21.5$ & $-4 - 5$ & 0.12$-$1.00 & 26\%\\
               & 18$^*$       & 37.5 (14.9) & 13.2 (4.0) & $-20.2$ (0.8)  & 1.0 (2.5) & 0.60 (0.20) \\
               &              & 11.4$-$79.2 & 3.3$-$21.1 & $-19.0- -21.4$ & $-4 - 5$ & 0.12$-$0.93 \\
               & 23$^\dagger$ & 54.5 (38.9) & 16.5 (6.8) & $-20.3$ (0.5)  & 3.0 (3.0) & 0.74 (0.16) \\
               &              & 11.4$-$119.4 & 3.3$-$24.9 & $-19.0- -21.5$ & $-4 - 5$ & 0.36$-$1.00 \\
\enddata 
\tablenotetext{a}{No superscript on the number corresponds
to ``matched samples'' (MS), $^*$~to the set with distance $\leq$ 80\,Mpc (DMS), and 
$^\dagger$~to the set with inclination $<$ 70$^\circ$ (IMS). }
\end{deluxetable} 
\clearpage

The most important parameter of relative spatial resolution,
parsec-to-pixel scale, is very similar for each of the final matched 
samples.
As seen in Table \ref{tab:medians}, the median resolutions are
10.4, 12.6, 7.5, 12.9, 15.2 pc/pixel for
the normal, \hii, LINER, Sy2, and Sy1 galaxies, respectively.
The worst discrepancy is a factor of two between 
LINERs and Sy1s, but as will be seen, this 
discrepancy only strengthens our results.

Distance is also another obviously important parameter, because of the 
inability to resolve features in far-away objects.
In the MS, the median distance of the Sy1s is 2.6 times larger
than the LINERs (the closest sample), and 30\% larger than
the Sy2s.
This effect is mitigated in the DMS, as the Sy1s are (in the median)
only twice as far as the LINERs, and at the same distance as the
Sy2s (see Table \ref{tab:medians}).

%


\begin{deluxetable}{lrrlrrrrrlllll} 
\rotate
\tablecolumns{14} 
\tableheadfrac{0.05} 
\tabletypesize{\scriptsize}
\tablewidth{0pt}
\tablenum{2}
\tablecaption{Sample Properties \label{tab:sample}}
\tablehead{ 
\multicolumn{1}{c}{Name\tablenotemark{a}} & 
\multicolumn{1}{c}{$z$} & \multicolumn{1}{c}{Dist.} &
\multicolumn{1}{c}{RC3 Type} &  \multicolumn{1}{c}{a} &
\multicolumn{1}{c}{b} &  \multicolumn{1}{c}{Mag.} &  \multicolumn{1}{c}{Abs. Mag.} &
\multicolumn{1}{c}{Pc/pixel} & \multicolumn{1}{c}{Bar} & 
\multicolumn{1}{c}{Twist} & \multicolumn{1}{c}{3$\theta$} & 
\multicolumn{1}{c}{cos(4$\theta$)} & \multicolumn{1}{c}{USM} \\
\colhead{(1)} &  \colhead{(2)} & 
\colhead{(3)} & \colhead{(4)} &
\colhead{(5)} & \colhead{(6)} &
\colhead{(7)} & \colhead{(8)} & 
\colhead{(9)} & \colhead{(10)} & 
\colhead{(11)} & \colhead{(12)} & 
\colhead{(13)} & \colhead{(14)} } 
\startdata
\cutinhead{Non-active}
NGC0289              & 0.005 &  18.5 & SAB(rs)bc       &  5.1 &  3.6 &  11.7 & $-19.6$  &  6.7 & Y? & Y? &    &    &    \\
NGC0488              & 0.008 &  27.5 & SA(r)b          &  5.2 &  3.9 &  11.2 & $-21.0$  & 10.0 &    &    &    &    &    \\
NGC0772              & 0.008 &  30.8 & SA(s)b          &  7.2 &  4.3 &  11.1 & $-21.4$  & 11.2 &    &    &    &    &    \\
NGC2196              & 0.008 &  31.7 & (R':)SA(rs)ab   &  2.8 &  2.2 &  11.8 & $-20.7$  & 11.5 &    &    &    &    &    \\
NGC2339              & 0.007 &  31.6 & SAB(rs)bc       &  2.7 &  2.0 &  12.5 & $-20.0$  & 11.5 &    &    & Y? & D? &    \\
NGC2460              & 0.005 &  21.5 & SA(s)a          &  2.5 &  1.9 &  12.7 & $-19.0$  &  7.8 & Y? &    &    &    &    \\
NGC2566              & 0.005 &  24.2 & (R')SB(r)ab     &  4.0 &  2.9 &  11.8 & $-20.1$  &  8.8 & Y? & Y? &    &    &    \\
NGC2748              & 0.005 &  21.9 & SAbc            &  3.0 &  1.1 &  12.4 & $-19.3$  &  8.0 &    &    &    & B? &    \\
ESO498-G005          & 0.008 &  35.9 & SAB(s)c         &  1.3 &  1.1 &  14.0 & $-18.8$  & 13.1 &    &    &    &    &    \\
NGC3067              & 0.005 &  23.8 & SAB(s)ab?       &  2.5 &  0.9 &  12.8 & $-19.1$  &  8.7 &    &    & Y? & D? &    \\
NGC3115              & 0.002 &  13.4 & S0-             &  7.2 &  2.5 &   9.9 & $-20.7$  & 13.0 &    &    &    & D  & $+$ \\
NGC3259              & 0.006 &  25.7 & SAB(rs)bc:      &  2.2 &  1.2 &  12.9 & $-19.1$  &  9.3 & Y? &    & Y? & D? &    \\
NGC3277              & 0.005 &  23.3 & SA(r)ab         &  1.9 &  1.7 &  12.5 & $-19.3$  &  8.5 &    &    &    &    &    \\
NGC3455              & 0.004 &  19.3 & (R')SAB(rs)b    &  2.5 &  1.5 &  12.8 & $-18.6$  &  7.0 &    &    &    & D? &    \\
NGC3900              & 0.006 &  29.1 & SA(r)0+         &  3.2 &  1.7 &  12.2 & $-20.1$  & 10.6 &    &    &    &    &    \\
NGC3949              & 0.003 &  14.7 & SA(s)bc:        &  2.9 &  1.7 &  11.5 & $-19.3$  &  5.3 & Y  &    &    &    &    \\
NGC4026$^\dagger$    & 0.003 &  16.2 & S0              &  5.2 &  1.3 &  11.7 & $-19.4$  & 15.7 &    &    &    & D  &    \\
NGC4219$^\dagger$    & 0.007 &  30.2 & SA(s)bc         &  4.3 &  1.3 &  12.7 & $-19.7$  & 11.0 & Y? &    &    &    & $-$ \\
NGC4417              & 0.003 &  16.1 & SB0:sp          &  3.4 &  1.3 &  12.0 & $-19.0$  & 15.6 &    &    &    & BD &    \\
NGC4806              & 0.008 &  37.1 & SB(s)c?         &  1.2 &  1.0 &  13.4 & $-19.4$  & 13.5 & Y? &    &    &    &    \\
ESO443-G080          & 0.007 &  32.7 & SB(s)m          &  1.4 &  0.9 &  14.1 & $-18.5$  & 11.9 &    &    &    &    & $+$ \\
NGC5326              & 0.008 &  38.6 & SAa:            &  2.2 &  1.1 &  12.9 & $-20.0$  & 14.0 &    &    &    &    &    \\
NGC5389$^\dagger$    & 0.006 &  28.3 & SAB(r)0/a:?     &  3.5 &  1.0 &  12.9 & $-19.4$  & 10.3 & Y? &    &    & B? & $+$ \\
NGC5377              & 0.006 &  28.2 & (R)SB(s)a       &  3.7 &  2.1 &  12.2 & $-20.1$  & 10.3 & Y? &    &    &    &    \\
NGC5422$^\dagger$    & 0.006 &  27.7 & S0              &  3.9 &  0.7 &  12.8 & $-19.4$  & 10.1 &    &    &    & B?D? &    \\
NGC5443              & 0.006 &  27.9 & SB(s)b?         &  2.7 &  1.0 &  13.1 & $-19.1$  & 10.1 & Y? &    &    & D? &    \\
NGC5448              & 0.007 &  31.3 & (R)SAB(r)a      &  4.0 &  1.8 &  11.9 & $-20.6$  & 11.4 &    & Y? &    &    &    \\
NGC5475$^\dagger$    & 0.005 &  25.7 & Sa?sp           &  2.0 &  0.5 &  13.5 & $-18.6$  &  9.3 &    &    &    & B? &    \\
IC4390$^\dagger$     & 0.007 &  31.3 & SA(s)b          &  1.8 &  0.6 &  13.8 & $-18.7$  & 11.4 & Y? &    & Y  &    &    \\
NGC5587$^\dagger$    & 0.008 &  35.9 & S0/a            &  2.6 &  0.8 &  13.5 & $-19.3$  & 13.1 &    &    &    & B?D? &    \\
NGC5689$^\dagger$    & 0.007 &  33.0 & SB(s)0/a:       &  3.5 &  1.0 &  12.8 & $-19.8$  & 12.0 & Y? &    & Y? & D? & $+$ \\
NGC5707$^\dagger$    & 0.007 &  33.6 & Sab:sp          &  2.6 &  0.4 &  13.3 & $-19.3$  & 12.2 & Y? &    &    & B?D? &    \\
NGC5719              & 0.006 &  27.8 & SAB(s)abpec     &  3.2 &  1.2 &  13.3 & $-18.9$  & 10.1 &    &    &    &    & $-$ \\
NGC5746$^\dagger$    & 0.006 &  27.5 & SAB(rs)b?sp     &  7.4 &  1.3 &  11.3 & $-20.9$  & 10.0 & Y? &    &    & D? & $-$ \\
NGC5806              & 0.005 &  22.3 & SAB(s)b         &  3.1 &  1.6 &  12.4 & $-19.3$  &  8.1 & Y? &    &    &    &    \\
NGC5854$^\dagger$    & 0.006 &  27.5 & SB(s)0+         &  2.8 &  0.8 &  12.7 & $-19.5$  & 10.0 &    &    &    & B  &    \\
NGC5965$^\dagger$    & 0.011 &  49.7 & Sb              &  6.2 &  0.8 &  12.6 & $-20.9$  & 18.1 & Y? &    &    & B?D? & $+$ \\
NGC6010$^\dagger$    & 0.006 &  28.9 & S0/a:sp         &  1.9 &  0.5 &  13.6 & $-18.7$  & 10.5 &    &    &    & D? &    \\
NGC6504$^\dagger$    & 0.016 &  67.4 & S               &  2.2 &  0.5 &  13.5 & $-20.6$  & 24.5 & Y? &    &    &    &    \\
NGC6684              & 0.003 &  11.4 & (L)SB(r)0$^+$     &  4.0 &  2.6 &  11.3 & $-19.0$  & 11.1 &    & Y  &    & D  &    \\
ESO404-G003          & 0.008 &  29.8 & SB(r)bc         &  1.5 &  0.7 &  13.9 & $-18.5$  & 10.8 &    &    &    &    & $+$ \\
NGC7162              & 0.008 &  28.7 & (R')SA(r)bc     &  2.8 &  1.0 &  13.3 & $-19.0$  & 10.4 & Y? &    &    &    &    \\
NGC7280              & 0.006 &  22.3 & SAB(r)0+        &  2.2 &  1.5 &  13.0 & $-18.7$  &  8.1 &    & Y? &    & D? &    \\
NGC7421              & 0.006 &  21.8 & SB(r)bc         &  2.0 &  1.8 &  13.0 & $-18.7$  &  7.9 &    &    & Y? & D? &    \\
IC5271               & 0.006 &  20.4 & Sb?             &  2.6 &  0.9 &  12.9 & $-18.6$  &  7.4 &    &    & Y? &    &    \\
IC5273               & 0.004 &  14.4 & SB(rs)cd        &  2.7 &  1.8 &  12.2 & $-18.6$  &  5.2 &    &    &    &    & $+$ \\
NGC7537$^\dagger$    & 0.009 &  33.1 & SAbc:           &  2.2 &  0.6 &  13.9 & $-18.7$  & 12.0 &    &    &    & D? & $+$ \\
\cutinhead{HII}
UGC01385             & 0.019 &  75.3 & (R)SB0/a        &  0.7 &  0.6 &  13.9 & $-20.5$  & 15.7 &    &    &    &    &    \\
NGC0986              & 0.007 &  24.5 & (R'$_1$)SB(rs)b &  3.9 &  3.0 &  12.0 & $-19.9$  &  8.9 &    &    &    &    & $+$ \\
NGC0972              & 0.005 &  18.6 & Sab             &  3.3 &  1.7 &  12.3 & $-19.1$  &  6.8 &    & Y? & Y? &    & $-$ \\
NGC2903              & 0.002 &  10.9 & SB(s)d          & 12.6 &  6.0 &   9.7 & $-20.5$  &  4.0 &    &    &    &    & $+$ \\
NGC2964              & 0.004 &  21.6 & SAB(r)bc        &  2.9 &  1.6 &  12.0 & $-19.7$  &  7.9 &    & Y? & Y? &    &    \\
NGC3184              & 0.002 &  11.5 & SAB(rs)cd       &  7.4 &  6.9 &  10.4 & $-19.9$  & 11.2 &    & Y  &    &    &    \\
NGC4062              & 0.003 &  14.7 & SA(s)c          &  4.1 &  1.7 &  11.9 & $-18.9$  & 14.3 &    & Y  &    &    &    \\
NGC4384              & 0.008 &  38.1 & Sa              &  1.3 &  1.0 &  13.5 & $-19.4$  & 13.9 &    &    & Y? & B? & $+$ \\
NGC4536              & 0.006 &  29.5 & SAB(rs)bc       &  7.6 &  3.2 &  11.2 & $-21.1$  & 10.7 & Y? &    & Y  &    & $+$ \\
NGC5188              & 0.008 &  36.6 & (R':)SAB(rs)b   &  3.0 &  1.1 &  13.0 & $-19.8$  & 13.3 & Y? &    & Y? & D? & $+$ \\
NGC5597              & 0.009 &  40.6 & SAB(s)cd        &  2.1 &  1.7 &  12.6 & $-20.4$  & 14.8 &    & Y? &    & B  &    \\
NGC5757              & 0.009 &  39.6 & (R')SB(r)b      &  2.0 &  1.6 &  13.5 & $-19.5$  & 14.4 &    &    &    &    & $+$ \\
NGC6000              & 0.007 &  32.5 & SB(s)bc:        &  1.9 &  1.6 &  13.0 & $-19.6$  & 11.8 &    &    &    &    & $+$ \\
NGC6207              & 0.003 &  13.8 & SA(s)c          &  3.0 &  1.3 &  12.2 & $-18.5$  & 13.4 &    &    & Y  &    &    \\
\cutinhead{LINER}
NGC1961              & 0.013 &  55.1 & SAB(rs)c        &  4.6 &  3.0 &  11.7 & $-22.0$  & 20.0 &    & Y  &    & D  &    \\
NGC2985              & 0.004 &  20.1 & (R')SA(rs)ab    &  4.6 &  3.6 &  11.2 & $-20.3$  &  7.3 &    &    &    &    &    \\
NGC3169              & 0.004 &  20.8 & SA(s)apec       &  4.4 &  2.8 &  11.1 & $-20.5$  &  7.6 &    &    &    &    &    \\
MESSIER105           & 0.003 &  16.6 & E1              &  5.4 &  4.8 &  10.2 & $-20.9$  & 16.1 &    &    &    &    &    \\
NGC3675              & 0.003 &  14.3 & SA(s)b          &  5.9 &  3.1 &  11.0 & $-19.8$  & 13.9 &    & Y? &    &    &    \\
NGC3898              & 0.004 &  19.4 & SA(s)ab         &  4.4 &  2.6 &  11.6 & $-19.8$  &  7.1 &    &    &    &    &    \\
NGC4102              & 0.003 &  14.9 & SAB(s)b?        &  3.0 &  1.7 &  12.0 & $-18.9$  &  5.4 &    &    &    &    & $-$ \\
NGC4143              & 0.003 &  17.4 & SAB(s)0$^0$      &  2.3 &  1.4 &  11.7 & $-19.5$  &  6.3 &    & Y? &    &    &    \\
NGC4293              & 0.003 &  16.7 & (R)SB(s)0/a     &  5.6 &  2.6 &  11.3 & $-19.8$  & 16.2 & Y? & Y  &    & B  &    \\
NGC4314              & 0.003 &  17.5 & SB(rs)a         &  4.2 &  3.7 &  11.4 & $-19.8$  &  6.4 &    & Y  &    & D  &    \\
NGC4527$^\dagger$    & 0.006 &  28.5 & SAB(s)bc/       &  6.2 &  2.1 &  11.4 & $-20.9$  & 10.4 & Y  &    & Y  &    & $-$ \\
NGC4750              & 0.005 &  24.6 & (R)SA(rs)ab     &  2.0 &  1.9 &  12.1 & $-19.9$  &  8.9 & Y? & Y? &    & D? &    \\
NGC5064              & 0.010 &  43.6 & (R':)SA(s)ab    &  2.5 &  1.1 &  12.9 & $-20.3$  & 15.9 &    &    & Y  &    &    \\
NGC5678              & 0.006 &  29.3 & SAB(rs)b        &  3.3 &  1.6 &  12.1 & $-20.2$  & 10.7 & Y? &    &    & B?D? &    \\
NGC5838              & 0.005 &  22.3 & SA0-            &  4.2 &  1.5 &  11.9 & $-19.8$  &  8.1 & Y? &    &    & D? &    \\
NGC5879$^\dagger$    & 0.003 &  13.2 & SA(rs)bc:?      &  4.2 &  1.3 &  12.2 & $-18.4$  &  4.8 &    &    &    & B? &    \\
NGC6340              & 0.004 &  17.8 & SA(s)0/a        &  3.2 &  3.0 &  11.9 & $-19.4$  &  6.5 &    &    &    &    &    \\
NGC6384              & 0.006 &  24.2 & SAB(r)bc        &  6.2 &  4.1 &  11.1 & $-20.8$  &  8.8 & Y? &    &    &    &    \\
NGC6744              & 0.003 &  10.6 & SAB(r)bc        & 20.0 & 12.9 &   9.1 & $-21.0$  &  3.9 &    &    &    &    &    \\
NGC7177              & 0.004 &  12.9 & SAB(r)b         &  3.1 &  2.0 &  12.0 & $-18.6$  &  4.7 &    & Y  & Y? & D? &    \\
NGC7217              & 0.003 &  10.6 & (R)SA(r)ab      &  3.9 &  3.2 &  11.0 & $-19.1$  &  3.9 &    &    &    &    &    \\
NGC7742              & 0.006 &  19.1 & SA(r)b          &  1.7 &  1.7 &  12.3 & $-19.1$  &  6.9 &    & Y? &    &    &    \\
\cutinhead{Seyfert 2}
NGC0449              & 0.016 &  63.3 & (R')S?          &  0.8 &  0.5 &  15.0 & $-19.0$  & 13.2 &    & Y? & Y  & D  &    \\
UGC01214             & 0.017 &  67.8 & (R)SAB(rs)0+:   &  1.3 &  1.3 &  13.7 & $-20.5$  & 24.7 & Y  & Y  &    &    &    \\
NGC0788              & 0.014 &  52.6 & SA(s)0/a:       &  1.9 &  1.4 &  13.0 & $-20.6$  & 19.1 &    & Y  &    &    &    \\
UGC02456             & 0.012 &  47.8 & (R)SB(s)0+      &  1.7 &  1.0 &  13.6 & $-19.8$  & 17.4 & Y? & Y  & Y  &    &    \\
NGC1241              & 0.014 &  52.9 & SB(rs)b         &  2.8 &  1.7 &  12.0 & $-21.6$  & 19.2 &    & Y  &    & B  &    \\
NGC1275              & 0.018 &  71.3 & cDpecNLRG       &  2.2 &  1.7 &  12.6 & $-21.7$  & 25.9 &    & Y  &    & B? &    \\
NGC1320$^\dagger$    & 0.009 &  34.6 & Sa:sp           &  1.9 &  0.6 &  13.3 & $-19.4$  & 12.6 &    &    &    & D  &    \\
NGC1398              & 0.005 &  16.6 & (R$_1$R'$_2$)SB(rs)ab &  7.1 &  5.4 &  10.6 & $-20.5$  &  6.0 &    &    & Y? &    &    \\
NGC1672              & 0.005 &  17.1 & (R'$_1$:)SB(r)bc &  6.6 &  5.5 &  10.3 & $-20.9$  &  6.2 &    & Y  &    &    &    \\
NGC1667              & 0.015 &  61.3 & SAB(r)c         &  1.8 &  1.4 &  12.8 & $-21.1$  & 22.3 &    & Y  &    &    &    \\
ESO362-G008          & 0.016 &  64.8 & Sa              &  1.2 &  0.6 &  13.6 & $-20.5$  & 13.5 &    & Y? &    & B  &    \\
UGC04203             & 0.013 &  58.0 & Sa              &  0.8 &  0.8 &  14.3 & $-19.5$  & 21.1 &    & Y  & Y? &    &    \\
NGC2681              & 0.002 &  11.8 & (R')SAB(rs)0/a  &  3.6 &  3.3 &  11.1 & $-19.3$  & 11.4 & Y? & Y  &    & D? &    \\
NGC2685              & 0.003 &  14.3 & (R)SB0+pec      &  4.5 &  2.3 &  12.1 & $-18.7$  & 13.9 &    &    &    & D  &    \\
NGC3081              & 0.008 &  36.0 & (R$_1$)SAB(r)0/a &  2.1 &  1.6 &  12.8 & $-20.0$  & 13.1 & Y  & Y  &    &    &    \\
NGC3079$^\dagger$    & 0.004 &  18.2 & SB(s)c          &  7.9 &  1.4 &  11.5 & $-19.8$  &  6.6 &    & Y  &    &    & $+$ \\
IC2560               & 0.010 &  43.2 & (R':)SB(r)bc    &  3.2 &  2.0 &  12.5 & $-20.7$  & 15.7 &    &    &    & B  &    \\
MESSIER096           & 0.003 &  16.4 & SAB(rs)ab       &  7.6 &  5.2 &  10.1 & $-21.0$  &  6.0 &    & Y  &    &    &    \\
NGC3486              & 0.002 &  13.4 & SAB(r)c         &  7.1 &  5.2 &  11.1 & $-19.5$  &  4.9 &    &    &    &    &    \\
NGC3593              & 0.002 &  12.9 & SA(s)0/a        &  5.2 &  1.9 &  11.9 & $-18.6$  & 12.5 &    &    & Y  &    &    \\
MESSIER066           & 0.002 &  14.3 & SAB(s)b         &  9.1 &  4.2 &   9.7 & $-21.1$  &  5.2 &    &    &    &    & $-$ \\
NGC3982              & 0.004 &  18.5 & SAB(r)b:        &  2.3 &  2.0 &  11.8 & $-19.5$  &  6.7 &    & Y  &    &    &    \\
NGC4388$^\dagger$    & 0.008 &  39.4 & SA(s)b:sp       &  5.6 &  1.3 &  11.8 & $-21.2$  &  8.2 &    & Y  &    &    & $-$ \\
MESSIER090           & -0.001 &  16.8 & SAB(rs)ab       &  9.5 &  4.4 &  10.3 & $-20.8$  &  6.1 & Y? &    &    &    &    \\
NGC4785              & 0.012 &  54.1 & (R')SAB(r)ab    &  1.9 &  1.0 &  13.2 & $-20.5$  & 19.7 &    & Y? &    &    &    \\
NGC4941              & 0.004 &  19.6 & (R)SAB(r)ab:    &  3.6 &  1.9 &  12.4 & $-19.1$  &  7.1 & Y  & Y  &    &    &    \\
NGC4939              & 0.010 &  47.3 & SA(s)bc         &  5.5 &  2.8 &  11.9 & $-21.5$  & 17.2 &    &    &    & B  &    \\
NGC4968              & 0.010 &  44.6 & (R')SAB0$^0$     &  1.9 &  0.9 &  13.9 & $-19.4$  & 16.2 &    & Y  &    & B  &    \\
NGC5005              & 0.003 &  17.0 & SAB(rs)bc       &  5.8 &  2.8 &  10.6 & $-20.6$  &  6.2 &    & Y? &    &    & $-$ \\
NGC5135              & 0.014 &  60.3 & SB(l)ab         &  2.6 &  1.8 &  12.9 & $-21.0$  & 21.9 &    &    &    &    & $+$ \\
NGC5256$^*$          & 0.027 & 117.4 & Compactpec      &  0.4 &  0.3 &  14.1 & $-21.2$  & 24.5 & Y? & Y  & Y  &    & $+$ \\
NGC5283              & 0.010 &  45.6 & S0?             &  1.1 &  1.0 &  14.2 & $-19.1$  &  9.5 & Y? & Y  &    & D? &    \\
UGC08718             & 0.016 &  71.3 & S               &  0.7 &  0.5 &  14.6 & $-19.7$  & 14.9 &    & Y  &    & B  &    \\
NGC5347              & 0.008 &  36.2 & (R')SB(rs)ab    &  1.7 &  1.3 &  13.4 & $-19.4$  & 13.2 &    & Y  & Y  &    &    \\
NGC5427              & 0.009 &  40.2 & SA(s)cpec       &  2.8 &  2.4 &  11.9 & $-21.1$  & 14.6 &    & Y  &    & D? &    \\
NGC5643              & 0.004 &  18.7 & SAB(rs)c        &  4.6 &  4.0 &  10.7 & $-20.7$  &  6.8 &    & Y  &    &    &    \\
NGC5695              & 0.014 &  62.1 & SBb             &  1.5 &  1.1 &  13.6 & $-20.4$  & 12.9 &    & Y  &    &    &    \\
NGC5929              & 0.008 &  37.4 & Sab:pec         &  1.0 &  0.9 &  14.1 & $-18.8$  &  7.8 &    & Y  &    & D  &    \\
NGC5953              & 0.007 &  30.4 & SAa:pec         &  1.6 &  1.3 &  13.3 & $-19.1$  & 11.1 &    & Y  &    &    &    \\
NGC6217              & 0.005 &  20.1 & (R)SB(rs)bc     &  3.0 &  2.5 &  11.8 & $-19.7$  &  7.3 & Y? & Y? &    &    &    \\
ESO137-G034          & 0.009 &  38.5 & SAB(s)0/a?      &  1.4 &  1.1 &  12.2 & $-20.7$  & 14.0 &    & Y  &    &    &    \\
ESO138-G001          & 0.009 &  38.3 & E-S0            &  1.0 &  0.5 &  14.3 & $-18.6$  & 13.9 & Y? & Y  &    &    &    \\
NGC6300              & 0.004 &  15.3 & SB(rs)b         &  4.5 &  3.0 &  11.0 & $-19.9$  &  5.6 &    &    &    &    &    \\
FAIRALL0049$^\dagger$$^*$ & 0.020 &  83.5 & Sa              &  0.0 &  0.0 &  13.2 & $-21.4$  & 17.4 &    &    &    & D  & $+$ \\
NGC6810$^\dagger$    & 0.007 &  26.8 & SA(s)ab:sp      &  3.2 &  0.9 &  12.4 & $-19.7$  &  9.7 & Y? &    & Y? & D? & $+$ \\
NGC6890              & 0.008 &  31.6 & (R')SA(r:)ab    &  1.5 &  1.2 &  13.0 & $-19.5$  & 11.5 &    & Y  & Y  &    &    \\
NGC6951              & 0.005 &  19.6 & SAB(rs)bc       &  3.9 &  3.2 &  11.6 & $-19.9$  &  7.1 &    & Y  &    &    & $-$ \\
IC5063               & 0.011 &  45.2 & SA(s)0+:        &  2.1 &  1.4 &  12.9 & $-20.4$  & 16.4 &    &    & Y  &    &    \\
NGC7130              & 0.016 &  64.1 & Sapec           &  1.5 &  1.4 &  13.0 & $-21.0$  & 23.3 & Y  & Y  &    &    &    \\
ESO075-G041$^*$      & 0.028 & 115.8 & SA0-Radiogal    &  1.6 &  0.8 &  14.3 & $-21.0$  & 24.1 &    &    &    &    &    \\
NGC7479              & 0.008 &  29.3 & SB(s)c          &  4.1 &  3.1 &  11.6 & $-20.7$  & 10.7 & Y? &    &    &    &    \\
NGC7496              & 0.006 &  19.4 & (R':)SB(rs)bc   &  3.3 &  3.0 &  11.9 & $-19.5$  &  7.1 &    & Y? & Y  & B  &    \\
NGC7582              & 0.005 &  18.3 & (R'$_1$)SB(s)ab &  5.0 &  2.1 &  11.4 & $-19.9$  &  6.7 &    &    &    &    & $+$ \\
NGC7674$^*$          & 0.029 & 116.4 & SA(r)bcpec      &  1.1 &  1.0 &  13.9 & $-21.4$  & 24.3 &    & Y? &    & D  &    \\
NGC7743              & 0.006 &  19.7 & (R)SB(s)0+      &  3.0 &  2.6 &  12.4 & $-19.1$  &  7.2 &    & Y  &    &    &    \\
\cutinhead{Seyfert 1}
UGC00006$^*$         & 0.022 &  87.7 & Pec             &  1.0 &  0.7 &  14.4 & $-20.3$  & 18.3 &    & Y  &    &    & $+$ \\
UGC01395             & 0.017 &  68.4 & SA(rs)b         &  1.3 &  1.0 &  14.2 & $-20.0$  & 14.3 &    &    &    &    &    \\
NGC1019$^*$          & 0.024 &  97.1 & SB(rs)bc        &  1.0 &  0.9 &  14.3 & $-20.6$  & 20.2 &    &    &    &    & $+$ \\
NGC1365              & 0.005 &  19.8 & (R')SBb(s)b     & 11.2 &  6.2 &  10.3 & $-21.2$  &  7.2 &    &    &    &    & $+$ \\
IC0450               & 0.019 &  79.2 & SAB0+:          &  0.8 &  0.5 &  15.0 & $-19.5$  & 16.5 &    &    &    &    & $+$ \\
NGC2639              & 0.011 &  48.5 & (R)SA(r)a:?     &  1.8 &  1.1 &  12.6 & $-20.8$  & 17.6 &    &    &    &    &    \\
NGC2841              & 0.002 &  11.4 & SA(r)b:         &  8.1 &  3.5 &  10.1 & $-20.2$  &  4.1 &    &    &    & B? &    \\
UGC05849$^*$         & 0.026 & 112.3 & Sc/d            &  0.9 &  0.6 &  14.7 & $-20.6$  & 23.4 &    & Y  & Y  &    & $+$ \\
NGC3516              & 0.009 &  38.8 & (R)SB(s)0$^0$:   &  1.7 &  1.3 &  12.5 & $-20.4$  & 14.1 &    & Y  &    & D  &    \\
NGC3786              & 0.009 &  41.2 & SAB(rs)apec     &  2.2 &  1.3 &  13.5 & $-19.6$  &  8.6 &    & Y  &    &    & $+$ \\
NGC4235$^\dagger$    & 0.008 &  37.8 & SA(s)a          &  4.2 &  0.9 &  12.6 & $-20.3$  & 13.7 &    &    &    & BD &    \\
NGC4253              & 0.013 &  58.0 & (R')SB(s)a:     &  1.0 &  0.8 &  13.7 & $-20.1$  & 21.1 &    &    &    & D  &    \\
NGC4278              & 0.002 &  13.2 & E1-2            &  4.1 &  3.8 &  11.2 & $-19.4$  & 12.8 &    &    &    &    &    \\
NGC4565$^\dagger$    & 0.004 &  22.0 & SA(s)b?sp3      & 15.9 &  1.9 &  10.4 & $-21.3$  &  8.0 &    &    &    &    & $-$ \\
NGC4593              & 0.009 &  41.7 & (R)SB(rs)b      &  3.9 &  2.9 &  11.7 & $-21.4$  & 15.2 &    &    &    &    & $+$ \\
NGC5033              & 0.003 &  16.0 & SA(s)c          & 10.7 &  5.0 &  10.8 & $-20.2$  &  3.3 &    & Y  & Y  & B  &    \\
NGC5252$^*$          & 0.023 & 100.2 & S0              &  1.4 &  0.9 &  14.0 & $-21.0$  & 20.9 &    &    &    & D? &    \\
NGC5273              & 0.004 &  18.8 & SA(s)0$^0$       &  2.8 &  2.5 &  12.4 & $-19.0$  &  3.9 &    & Y  &    &    &    \\
NGC5506$^\dagger$    & 0.006 &  29.6 & Sapecsp         &  2.8 &  0.9 &  13.4 & $-18.9$  & 10.8 &    &    &    & B  &    \\
NGC5674$^*$          & 0.025 & 107.5 & SABc            &  1.1 &  1.0 &  13.7 & $-21.5$  & 22.4 &    & Y? & Y  & D  &    \\
NGC5985              & 0.008 &  37.1 & SAB(r)b         &  5.5 &  3.0 &  11.9 & $-20.9$  & 13.5 & Y? &    &    & D? &    \\
NGC6104$^*$          & 0.028 & 119.4 & S(R)pec/Pec     &  0.8 &  0.7 &  14.2 & $-21.2$  & 24.9 &    & Y  &    &    &    \\
ESO103-G035          & 0.013 &  54.5 & SA0$^0$           &  1.1 &  0.4 &  14.7 & $-19.0$  & 19.8 &    &    &    &    &    \\
NGC6814              & 0.005 &  20.2 & SAB(rs)bc       &  3.0 &  2.8 &  12.1 & $-19.4$  &  7.3 &    & Y? &    & D? &    \\
MRK0516$^*$          & 0.028 & 115.0 & Sc              &  0.5 &  0.5 &  15.3 & $-20.0$  & 24.0 &    &    & Y  & D  &    \\
MRK0915$^\dagger$$^*$ & 0.024 &  96.5 & Sb              &  1.0 &  0.3 &  14.8 & $-20.1$  & 20.1 &    &    &    & B  &    \\
UGC12138$^*$         & 0.025 & 100.3 & SBa             &  0.8 &  0.7 &  14.2 & $-20.8$  & 20.9 &    &    &    & D  &    \\
\enddata 
\tablenotetext{a}{
$^*$~Eliminated in the DMS because distance $>$ 80\,Mpc. 
$^\dagger$~Eliminated in the IMS because inclination $>$ 70$^\circ$. 
} 
\end{deluxetable} 
\clearpage

The images ($\S$\ref{sec:reduction}), 
surface brightness profiles with ellipse parameters ($\S$\ref{sec:ellipse}), and
unsharp-masked images ($\S$\ref{sec:usm}) are shown in Figures \ref{fig:imgnrm},
\ref{fig:imghii}, \ref{fig:imglin}, \ref{fig:imgsy2}, and
\ref{fig:imgsy1}, for the normal, \hii, LINER, Sy2, and Sy1 samples, respectively.
Only a representative page of each activity sample is shown; the remainder 
are available electronically (\url{www.arcetri.astro.it/$_{\tilde{\,}}$hunt/nicmos.html}).

\placefigure{fig:imgnrm}
\placefigure{fig:imghii}
\placefigure{fig:imglin}
\placefigure{fig:imgsy2}
\placefigure{fig:imgsy1}

\section{The Analysis \label{sec:analysis}}

We have studied the circumnuclear morphology of these galaxies with several
methods.  First, 
elliptically-averaged profiles were generated. Then, the
elliptical surface brightness distribution was subtracted from 
and divided by the original image to create unsharp masks (USMs); 
with these, we are better able to examine
residual asymmetric structures not well fitted with ellipses.  Finally,
objective quantitative procedures were applied to identify all morphological
peculiarities discussed here, with subsequent visual inspection to verify the
objective diagnostic. The morphological peculiarities we examined are outlined
below, together with their operational definition and what kinds of physical
processes they probe.

\subsection{Elliptical Isophote Fitting \label{sec:ellipse}}

We fitted an axisymmetric Gaussian to the nuclear region of each galaxy to 
determine the centers.  Then, we used the IRAF/STSDAS task 
{\it isophote/ellipse}\footnote{STSDAS is distributed by the Space
Telescope Science Institute, which is operated by the Association of
Universities for Research in Astronomy (AURA), Inc., under NASA contract
NAS 5--26555.} to fit the major and minor axes, position angle and brightness
level of a series of elliptical isophotes, logarithmically spaced in 
galactocentric distance.
Coefficients to the cos 3$\theta$ and cos 4$\theta$ residual
terms were also determined \citep{jed}.
Except for the ellipse center, which was kept fixed,
all the coefficients were allowed to vary over the full radial range.
We also generated a set of profiles with linear spacing.
Both types of profiles were analyzed as described in $\S$\ref{sec:morph}.

The elliptically-averaged surface-brightness profiles together with
the higher-order residuals
are plotted as a function of radius in the central panels of
Figures \ref{fig:imgnrm}--\ref{fig:imgsy1}.
Only the logarithmically spaced profiles are shown in the Figures.
Plots for all the galaxies in the sample are available
electronically from 
\url{www.arcetri.astro.it/$_{\tilde{\,}}$hunt/nicmos.html}.

A few profiles show an apparent small inflection in the surface brightness
at a radius of 0.5\arcsec\ (e.g., NGC\,5443, NGC\,5475, NGC\,5587,
NGC\,5854, NGC\,4293, NGC\,5838).
This is an artifact due to the ellipse fitting algorithm which we set to start
at a 0.5\arcsec\ radius with an initial value of 0 for the ellipticity.
In these galaxies,
the algorithm was able to fit the central region within 0.5\arcsec\
only with circular isophotes; beyond this, the best fit was obtained with
elliptical isophotes.
 
\subsection{Unsharp Masking from Ellipse Fitting \label{sec:usm}}

Using the STSDAS task {\it bmodel}, we converted the fitted elliptical isophotes into
a smooth model of the galaxy surface brightness distribution, and subtracted it from 
the original 
image, out to a typical radius of typically 100 pixels.
The resulting residual images are then normalized by the original image.
Examples are shown in the right panels of 
Figures \ref{fig:imgnrm}--\ref{fig:imgsy1}, which present the 
fractional deviations of the brightness distributions from the purely elliptical 
fitted isophotes. 
These images, very much like unsharp masks, filter out the low
spatial frequencies, and 
show the fine-structure residual structures that cannot
be fitted by any smooth symmetric model.
Since we kept the center of the concentric ellipses fixed, any central structure
is due to non-axisymmetric structure on small spatial scales.
A comparison of our unsharp mask images with those 
in common with \citet{swara} shows that they are virtually identical.
Although our methods differ, the results are similar because
no model with only axisymmetric components, even if fully two-dimensional, can
fit these fine structures.


 
\subsection{Reliability Checks}

11 galaxies in our sample were observed more than once, by different
NICMOS cameras or different observers (e.g., UM\,146, NGC\,5033, NGC\,5252, and NGC\,5273
with NIC-1 and NIC-2; NGC\,1241, NGC\,2639, NGC\,2841, and NGC\,3627,
NGC\,4102, and NGC\,6744 with NIC-2 and NIC-3; NGC\,2985 by different
observers with NIC-2).
Although these frames were reduced independently, we find that
the resulting photometry, isophotes and visual appearances are virtually
identical, as shown in Figure \ref{fig:comp}.

\placefigure{fig:comp}

The only disagreement larger than a few hundredths of a magnitude
is for the nuclear region of NGC\,5033;
however the two profiles are identical beyond a radius of 0.15\arcsec.
The position angles ($\theta$) may disagree because of the different
orientations of the observations, since the profile extraction and analysis
was performed on the original (unrotated to canonical North up, East left)
images. 
The generally excellent agreement gives us confidence that the images, 
photometry and surface brightness
profiles analyzed below are accurate and reproducible.
This also means that $H$-band nuclear variability above 10-20\% is not
very common, as already mentioned in $\S$\ref{sec:phot}.
In all cases, we have incorporated only the higher resolution images
in the analysis.

A number of the galaxy images presented in this study have already been
reduced and analyzed independently by \citet{swara} and \citet{laine02}.
We have 14 galaxies in common with \citet{swara}. 
By comparing their brightness profiles and unsharp masks with ours, 
we find generally very good agreement.
Only in four cases do our masks not reveal the structure that they find
with two-dimensional bulge/disk decomposition models.
In all these, theirs show very faint {\it axisymmetric} features, while our USMs are 
featureless.
This shows that our USMs are as efficient as more sophisticated ones
in revealing the small-scale {\it non-axisymmetric} structure
that we are interested in.
The profile parameters ($\epsilon$ and $\theta$) of the 67 galaxies in 
common with \citet{laine02} also agree well.
However, the agreement is worse when the galaxy is more inclined;
this is probably because \citet{laine02} deproject their profiles 
in order to analyze bar properties on all scales while we do not.
When the galaxies are face-on or almost, our profiles are identical to
theirs.

\subsection{Central Fine Structure \label{sec:morph}}

A ``profile analyzer'' was applied to the {\it linearly-spaced} profiles.
This objective algorithm  follows each profile and calculates the extrema
and radial variations of the fitted ellipticities, position angles, $3\theta$, 
and $\cos (4\theta)$ coefficients. 
Probable morphological features are identified automatically in each profile,
but the profiles were subsequently inspected visually by both
authors independently to ensure against spurious features.
The logarithmically-spaced profiles were also subjected to the analyzer,
then checked visually as before.
When a feature was clear in the linearly-spaced profiles, but less so in
the log ones, a ``?'' was assigned to it.
These relatively more uncertain features are given half-weight in the 
subsequent statistical comparison (see also $\S$\ref{sec:stat}).
All 250 profiles were analyzed before the galaxy samples were compiled;
in principle no bias was introduced because of preconceived
knowledge of activity type.
For each galaxy, our findings of
bars, isophotal twists, large $3\theta$ coefficients, boxy/disky isophotes,
and high-amplitude non-axisymmetric structure as identified in the unsharp masks
are reported in Table \ref{tab:sample}.

\subsubsection{Bars}

Bars are defined in the profile analyzer (and visually) according to
\citet{mcleod} and \citet{wozniak95}; the requirement is 
that the fitted position angle (PA or $\theta$) remains constant to within
$10^\circ$, while the ellipticity $\epsilon$ monotonically rises to a maximum, then
falls to some value, which on larger scales is usually determined by
the galaxy inclination  ($\cos(i)=b/a$). 
This is a slightly different definition of a bar than that of \citet{laine02},
who require that the PA remains constant to within $20^\circ$.
Indeed, some of the features that we call isophotal twists (see below) may be identified 
as bars in \citet{laine02}, a point which will be discussed in more detail in 
$\S$ \ref{sec:stat}.

Imaging in the NIR is particularly sensitive to stellar bars.
Bars imply non-axisymmetric radial 
streaming motions (e.g., \citealt{binneytremaine}) which may be relevant
to nuclear fueling or a massive compact central object.  

\subsubsection{Isophotal Twists}

Following \citet{wozniak95}, \citet{elmegreen96}, and \citet{jungwiert}, 
isophotal twists are defined   as systematic rotations in fitted PA $\theta>10^\circ$
over a region with monotonically varying ellipticity $\epsilon$. 
When twists occurred over the same radial range as the effects of a strong
nuclear point-spread function (PSF), they were not considered significant.

Isophotal twists may be related to stellar orbits and resonances 
or triaxial structure \citep{shaw93}, although
the presence of a resonance does not guarantee a twist \citep{elmegreen96}.
Twists may also be related to nested bars (e.g., \citealt{shlosman89}),
either through gas viscosity and dissipation \citep{shaw93} or
through two misaligned bars at different pattern speeds \citep{friedlimartinet}.

\subsubsection{$3\theta$ and $4\theta$ Coefficients}

The higher-order ($3\theta$ and $4\theta$) residuals to best-fit ellipses can be 
the diagnostic of dynamical instabilities in the stellar component.
Significant boxy or disky isophotes are identified in those
profiles where the cos4$\theta$ coefficient, A4, is non-zero over a
substantial range in radius; in the case of disky profiles,
A4$>$0, and for boxy ones, A4$<$0 \citep{carter,jed}.
For a profile to obtain boxy or disky status,
the A4 coefficient must be $>0.02$ (these are normalized, see documentation
of the IRAF/STSDAS {\it ellipse} task) over a range of radius well outside
the nuclear PSF.
This is because the NICMOS PSF tends to be boxy, and strong nuclei generally
showed negative A4 terms close to the nucleus.

We defined significant cos3$\theta$ or sin3$\theta$ residuals in profiles
where the coefficients of these terms, A3 or B3, are non-zero over
some range in radius, and larger than the cos4$\theta$ residuals
over the same range. 
Given the fitting procedure followed in {\it ellipse}, it is unlikely
that the same galaxy image can show both strong A3/B3 coefficients as well
as strong A4 coefficients (``boxiness") over the same radii.
We confirmed that these two classifications are virtually mutually
exclusive in our study: the presence of strongly detectable boxiness
eliminates the possibility of detecting strong A3/B3 asymmetry, and
vice-versa. 

Boxy/disky isophotes in the central regions of elliptical galaxies
have been investigated with numerical simulations, which suggest
that they may originate in mergers of disk galaxies \citep{naab}.
It is unclear whether this phenomenon could explain such
isophotes in the galaxies observed here, because they have 
retained their stellar disk.
Nevertheless, it may have bearing on 
the merger origin of Seyfert activity as proposed by
\citet{dultzin99,krongold}.
Boxy isophotes and $3\theta$ excesses in disk galaxies may 
also be related to internal dynamical instabilities and vertical resonances 
\citep{merritthernquist,pfenniger,patsis}.
However,
strong A3/B3 coefficients primarily measure distortions from dust filaments
\citep{peletier90}, 
although they can show up as morphological disturbances produced by close
gravitational encounters with nearby companion galaxies \citep{kenney}.

\subsubsection{Unsharp-Masked Images} 

Each residual image was searched automatically for regions of particularly large
positive or negative deviations from the model fit.
A grid of squares was superposed onto each galaxy, with a length
set to the integral number of pixels closest to a physical distance of 180 parsecs.
(These squares were on average about 12 x 12 pixels in size).
This search was repeated with the grid of squares shifted by half a box in
both coordinates.
For each search and each box, the median, mean, and mode of the residual
image was calculated, and
the boxes with the highest and lowest median values were identified.
The nuclear region was avoided in the searches because of possible contamination 
by a strong nuclear PSF.
We defined a ``significant'' deviation in the unsharp-masked images
(normalized residuals from the smooth ellipse fit, hereafter USM) as one with
an absolute value of the median in one or boxes of 0.27 or greater.
This somewhat arbitrary cut-off was established by visual inspection
of all the USMs, and the consequent evaluation of what was a real feature. 
Bad pixels had been previously removed by the reduction algorithm, but
we checked to make sure that none of the significant USMs was defined so
because of bad pixel contamination.
The last column (14) of Table \ref{tab:sample} lists all USMs
which have any 180-parsec boxes which deviate from the smooth model fit
by more than +/- 27\%; those with positive deviations are designated
with $+$, and negative ones with $-$.

Negative USM residuals are typically associated with dust
(e.g., \citealt{sparks85}), while positive features may indicate star
clusters or compact \hii\ regions, similar to color images \citep{pogge02}. 
The unsharp mask structure with this technique is usually very similar to
emission-line images when these last are available (e.g., \citealt{boker}),
and in most cases also to $V-H$ color images \citep{martini99,martini03a}.
Good examples of this agreement are NGC\,3786, UGC\,12138, NGC\,5033,
NGC\,5252, NGC\,5273 (Sy1s), and NGC\,5347, NGC\,5929, and NGC\,7674 (Sy2s).
Because negative USM residuals tend to indicate the presence of dust, it is 
important to compare the $3\theta$ diagnostic with the USM one.

\section{Results: Comparison of Seyfert, LINER, HII and Normal Galaxies \label{sec:stat}}

\begin{figure*}[t]
\centering
{\includegraphics[width=0.48\linewidth, bb=125 140 545 760]{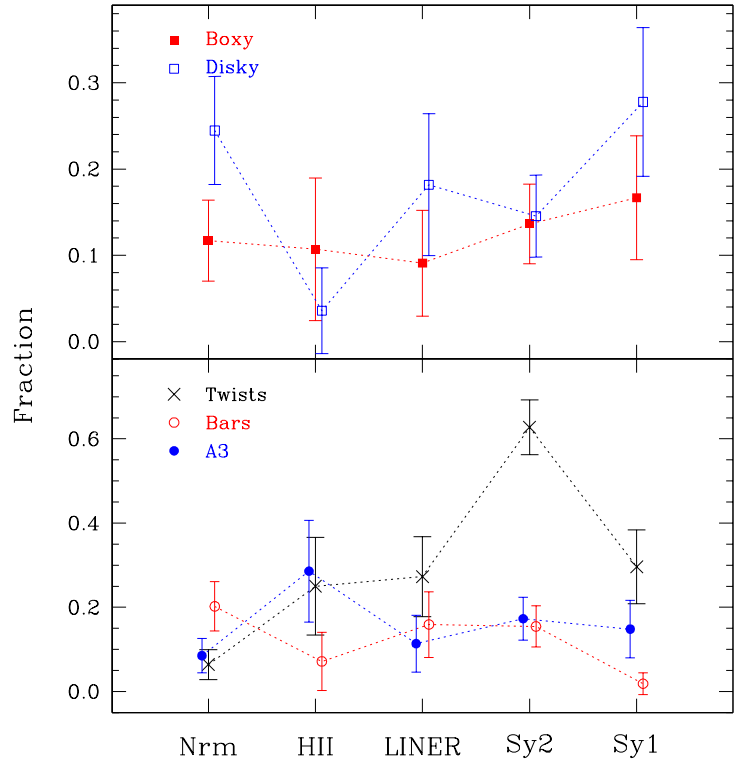}}
\caption{Fractions of non-axisymmetric features as a function
of activity type.
The error bars are calculated as $\sigma\,=\,\sqrt{N_{act} (1-N_{act})/T_{act}}$,
where $N_{act}$ is the number of features in activity class {\it act},
and $T_{act}$ is the total number in the class.
Individual features are joined only for the eye.
\label{fig:nonaxy}}
\end{figure*}

\begin{figure*}[t]
\centering
{\includegraphics[width=0.70\linewidth, bb=20 230 590 600]{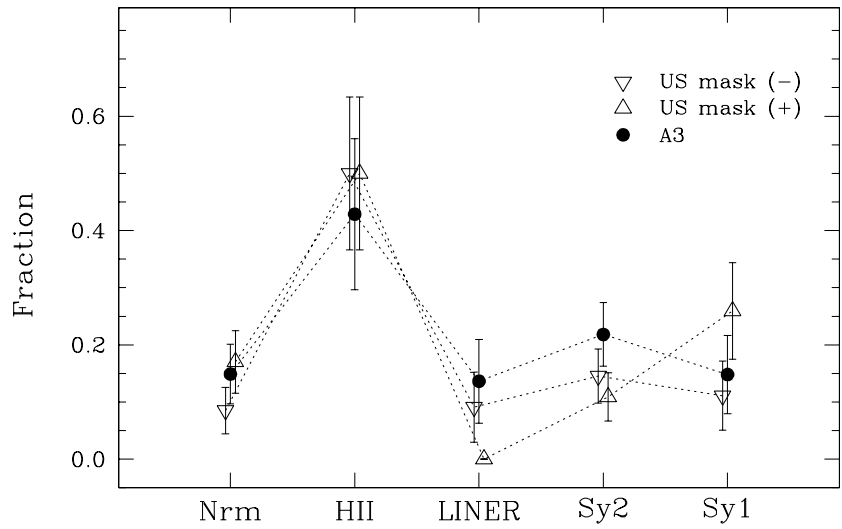}}
\caption{Fractions of positive (upward-pointing triangle)
and negative (downward triangle) USM residuals as a function
of activity type.
The error bars are calculated as in Fig. \ref{fig:nonaxy}.
Individual features are joined only for the eye.
The A3/B3 comparison shown here differs from that in Fig. \ref{fig:nonaxy}
because {\it full} (rather than half) weighting is given to the 
uncertain classifications.
\label{fig:usm}}
\end{figure*}

The fractions of each morphological diagnostic as a function of activity class 
(normal, HII/starburst, LINER, Sy2, Sy1) are reported in Table \ref{tab:features},
and shown graphically in Figures \ref{fig:nonaxy} and \ref{fig:usm}.  
Fractions are calculated by assigning unit weight to ``certain'' identifications
(when the features were evident in both the linear- and logarithmically-spaced
profiles), and half weight to less certain ones.
Figure \ref{fig:nonaxy} shows the frequencies of small-scale bars, 
isophotal twists, boxy/disky isophotes, 
and the cos$3\theta$ and sin$3\theta$ residual (denoted as A3/B3).  
Figure \ref{fig:usm} shows these frequencies for the indicators of
strong nuclear asymmetry, 
namely the positive/negative residuals measured in the unsharp masks, 
and the cos$3\theta$ and sin$3\theta$ residuals (repeated from Fig. \ref{fig:nonaxy}). 

\placefigure{fig:nonaxy}
\placefigure{fig:usm}

\clearpage
\begin{deluxetable}{lccccccccc} 
\rotate
\tablecolumns{10} 
\tableheadfrac{0.05} 
\tabletypesize{\scriptsize}
\tablewidth{0pt}
\tablenum{3}
\tablecaption{Non-Axisymmetric Features: Matched Sample Fractions\tablenotemark{a} \label{tab:features}}
\tablehead{ 
\multicolumn{1}{c}{Activity} & 
\multicolumn{1}{c}{Sample\tablenotemark{b}} & 
\multicolumn{1}{c}{Number} & 
\multicolumn{1}{c}{Isophotal} &
\multicolumn{1}{c}{Bars} &
\multicolumn{1}{c}{$3\theta$} & 
\multicolumn{1}{c}{Boxy} & 
\multicolumn{1}{c}{Disky} &
\multicolumn{1}{c}{USM} &
\multicolumn{1}{c}{USM} \\ 
\multicolumn{1}{c}{Class} & & & 
\multicolumn{1}{c}{Twists} & & 
\multicolumn{1}{c}{A3/B3} & 
\multicolumn{1}{c}{A4$<$0} & 
\multicolumn{1}{c}{A4$>$0} &
\multicolumn{1}{c}{$<$0} &
\multicolumn{1}{c}{$>$0} \\ 
\colhead{(1)} &  \colhead{(2)} & 
\colhead{(3)} & \colhead{(4)} &
\colhead{(5)} & \colhead{(6)} & 
\colhead{(7)} & \colhead{(8)} & 
\colhead{(9)} & \colhead{(10)} }
\startdata
Normal & Matched (MS)         & 47 & 0.06\ \ 0.04 & 0.20\ \ 0.06 & 0.09\ \ 0.04 & 0.12\ \ 0.05 & 0.24\ \ 0.06 & 0.09\ \ 0.04 & 0.17\ \ 0.05\\
       & D$\leq$80\,Mpc (DMS) &  47 & 0.06\ \ 0.04 & 0.20\ \ 0.06 & 0.09\ \ 0.04 & 0.12\ \ 0.05 & 0.24\ \ 0.06 & 0.09\ \  0.04 &  0.17\ \  0.05 \\ 
       & $\cos(i)>0.34$ (IMS) & 32 & 0.09\ \ 0.05 & 0.17\ \ 0.07 & 0.08\ \ 0.05 & 0.05\ \ 0.04 & 0.20\ \ 0.07 & 0.03\ \ 0.03 & 0.13\ \ 0.06\\
\\
\hii/ & Matched (MS)          & 14 &0.25\ \ 0.12 & 0.07\ \ 0.07 & 0.29\ \ 0.12 & 0.11\ \ 0.08 & 0.04\ \ 0.05 & 0.50\ \ 0.13 & 0.50\ \ 0.13\\
      & D$\leq$80\,Mpc (DMS) &  14 & 0.25\ \ 0.12 & 0.07\ \ 0.07 & 0.29\ \ 0.12 & 0.11\ \ 0.08 & 0.04\ \ 0.05 &  0.50\ \  0.13 &  0.50\ \  0.13 \\
starburst & $\cos(i)>0.34$ (IMS) & 14 &0.25\ \ 0.12 & 0.07\ \ 0.07 & 0.29\ \ 0.12 & 0.11\ \ 0.08 & 0.04\ \ 0.05 & 0.50\ \ 0.13 & 0.50\ \ 0.13\\\
\\
LINER  & Matched (MS)         & 22 &0.27\ \ 0.09 & 0.16\ \ 0.08 & 0.11\ \ 0.07 & 0.09\ \ 0.06 & 0.18\ \ 0.08 & 0.09\ \ 0.06 & 0.00\ \ 0.00\\
       & D$\leq$80\,Mpc (DMS) &  22 & 0.27\ \ 0.09 & 0.16\ \ 0.08 & 0.11\ \ 0.07 & 0.09\ \ 0.06 & 0.18\ \ 0.08 &  0.09\ \  0.06 &  0.00\ \  0.00 \\ 
       & $\cos(i)>0.34$ (IMS) & 20 &0.30\ \ 0.10 & 0.12\ \ 0.07 & 0.07\ \ 0.06 & 0.07\ \ 0.06 & 0.20\ \ 0.09 & 0.05\ \ 0.05 & 0.00\ \ 0.00\\
\\
Sy 2   & Matched (MS)  & 55 &0.63\ \ 0.07 & 0.15\ \ 0.05 & 0.17\ \ 0.05 & 0.14\ \ 0.05 & 0.15\ \ 0.05 & 0.15\ \ 0.05 & 0.11\ \ 0.04\\
       & D$\leq$80\,Mpc (DMS) &  51 & 0.65\ \ 0.07 & 0.16\ \ 0.05 & 0.17\ \ 0.05 & 0.15\ \ 0.05 & 0.12\ \ 0.05 &  0.16\ \  0.05 &  0.08\ \  0.04 \\ 
       & $\cos(i)>0.34$ (IMS) & 50 &0.65\ \ 0.07 & 0.16\ \ 0.05 & 0.18\ \ 0.05 & 0.15\ \ 0.05 & 0.11\ \ 0.04 & 0.10\ \ 0.04 & 0.06\ \ 0.03\\
\\
Sy 1   & Matched (MS)         & 27 &0.30\ \ 0.09 & 0.02\ \ 0.03 & 0.15\ \ 0.07 & 0.17\ \ 0.07 & 0.28\ \ 0.09 & 0.11\ \ 0.06 & 0.26\ \ 0.08\\
       & D$\leq$80\,Mpc (DMS) & 18 & 0.25\ \ 0.10 & 0.03\ \ 0.04 & 0.06\ \ 0.05 & 0.19\ \ 0.09 & 0.22\ \ 0.10 &  0.17\ \  0.09 &  0.22\ \  0.10 \\   
       & $\cos(i)>0.34$ (IMS) & 23 &0.35\ \ 0.10 & 0.02\ \ 0.03 & 0.17\ \ 0.08 & 0.07\ \ 0.05 & 0.28\ \ 0.09 & 0.09\ \ 0.06 & 0.30\ \ 0.10\\
\enddata 
\tablenotetext{a}{Unit weight is given to ``certain'' features; half weight is given
to ``uncertain'' features which are less clear in the logarithmically spaced profiles
than in the linear ones (see text).}
\tablenotetext{b}{First line: matched samples; 
second line: matched samples with additional distance constraint;
third line: matched samples with inclination constraint.}
\end{deluxetable} 
\clearpage


To determine statistical significance,
we compared the structural properties among the
various activity sub-samples using the z-test 
\citep{mccabe}\footnote{This test requires a z score of greater
than 1.65 for differences which have a 5\% or lower probability of being
due to chance (95\% significance).}.
Uncertainties in the fractions are given as
$\sigma\,=\,\sqrt{N_{act} (1-N_{act})/T_{act}}$,
where $N_{act}$ is the number of features in activity class {\it act},
and $T_{act}$ is the total number in the class.
We list in the following {\it only those trends which are
$\geq$ 95\% significant in all three sample sets (MS, DMS, IMS)}:
\begin{itemize}
\item
There are more isophotal twists in Sy2s, and 
fewer of them in non-active galaxies than in any other class.
63$\pm$7\% of the Sy2s possess twisted isophotes, in contrast
with only 6$\pm$4\% the normal galaxies.
The significance levels of these differences range from 
2.2$\sigma$ (Sy2, non-active vs. \hii, LINER, Sy1) to 5.8$\sigma$ (Sy2 vs. non-active).
\item
Bars are less frequent in Sy1s relative to normal galaxies;
2$\pm$3\%\footnote{This corresponds to 0.5 galaxies, because of the half 
weighting used for uncertain determinations.} of the Sy1s 
and 20$\pm$6\% non-active galaxies
have nuclear bars according to our definition.
This difference is a 2$\sigma$ effect (98\% confidence level).
\item
29$\pm$12\% of the \hii\ galaxies have strong $3\theta$ residuals
compared to 9$\pm$4\% of non-active galaxies. 
This is a significant excess at a confidence level of 98\% (2$\sigma$).
\item
Only 4$\pm$5\% of \hii/starbursts contain disky isophotes,
a deficiency significant at a 2$\sigma$ level relative to the 
the 24$\pm$6\% fraction of non-active galaxies .
There is no difference among the samples for boxy isophotes.
\item
50$\pm$13\% of the \hii\ galaxies contain either positive or negative USM residuals.
{\it Negative} USM residuals occur more often in \hii/starbursts than in any other 
activity class, differences significant at $2.8-3.5\sigma$.
{\it Positive} USM residuals are also more frequent in \hii's than in 
non-active galaxies (2.5$\sigma$), 
LINERs (3.7$\sigma$), and Sy2s (3.3$\sigma$), and 
also more frequent in Sy1s relative to Sy2s (1.8$\sigma$) and 
LINERs (2.6$\sigma$).  
They are less frequent (0 objects) in LINERs than in any other class.  
\end{itemize}

In addition to the three main sets of samples (MS, DMS, IMS),
we also divided the samples into high- and low-luminosity groups,
and recalculated the statistics.
The significantly greater frequencies of isophotal twists in Sy2s, 
A3 and USM positive residuals in \hii/starbursts, and
fewer bars in Sy1s emerge as before.
For particular features (see $\S$\ref{sec:twists}, \ref{sec:3theta}, \ref{sec:usmresiduals}),
we also derived statistics on subsets of early and late Hubble types.
The significance of the trends was usually enhanced.
We therefore conclude that these results are robust to
possible sample biases, and in what follows,
each feature is discussed in detail.

\subsection{Isophotal Twists \label{sec:twists}}

More than 60\% of Sy2 galaxies show twisted isophotes, while
the frequency of twists in all of the remaining active samples is 
roughly 20$-$30\%.
Only the non-active galaxies show a very low fraction of 6\%.
It is difficult to attribute this result to
different sensitivity to structure
on a given spatial scale, because the median resolution of the Sy2
sample is very similar to the non-active and \hii\ samples (Table \ref{tab:medians}).

We have examined the possibility that the isophotal twists 
are related to primary large-scale bars (e.g., \citealt{shaw93};
\citealt{friedlietal}; \citealt{jungwiert}).
By considering the bar classes from RC3, and then
tallying the twisted isophotes which occur in barred galaxies, 
we find that the majority of them are found in the SAB and SB
galaxies, with the exception of the Sy1s.
However, while 3 of 3 twists (100\%)
in the non-active sample are associated with SB or SAB morphology,
and 75\% in LINERs,
the percentage decreases considerably for the remaining activity types.
Only 57\% of the twisted isophotes in \hii\ (2 of 3.5 twists)
and Sy2 galaxies (19.5 of 34.5) are found in barred galaxies,
and 38\% (3 of 8) of the twists in Sy1s.
Although these numbers suffer to some degree 
from small-number statistics, they indicate that a substantial
fraction of circumnuclear twists in Seyferts are not found in previously 
known barred galaxies.

We also investigated the possibility that the differences in
frequency of twisted isophotes among the samples 
are related to Hubble type, since they tend to be found
primarily among early spirals \citep{friedlietal,elmegreen96}.
If we divide the samples into late ($T\geq3$) and early ($T<3$) types
and redo the analysis, we find that
the same statistical differences are shown by
both the early and late sub-samples.
Indeed, the isophotal twists in our sample are
not confined to strictly early-type morphologies, as they
are seen in Hubble types as late as Sc (e.g., NGC\,5643). 
The early Hubble types taken alone show significantly more twists
in Sy2s relative to non-active galaxies, LINERs, and Type 1 
Seyferts as before\footnote{The small number of early types in the \hii\ 
sample are not not sufficient to make the differences significant, although
the same trend is present.}.
We conclude that the greater frequency of isophotal twists in Sy2s, 
and their lesser frequency in non-active galaxies,
are not caused by differences in Hubble types among the samples.

\subsection{Bars \label{sec:bars}}

Even though $\S$\ref{sec:bartypes} shows that 84\% of the \hii\ sample contains 
{\it large-scale}
bars and more than half the galaxies in the remaining samples are barred,
there are very few bars on the scales probed by our NICMOS images.
The largest bar fraction we find is $\sim20$\% in the non-active sample;
the active galaxies have fewer bars, $\sim15$\%, and 
the Sy1s show no bars at all (at best we have an uncertain determination,
which we have given 0.5 weighting in the analysis).
It is likely that the low bar fraction in Sy1s and starbursts is related
to the excess of positive USM residuals (see $\S$\ref{sec:usmresiduals});
both have significantly high fractions of these,
together with anomalously low fractions of bars.
Such irregular structure would make it difficult for the profile
analyzer to pick out a bar.
Also a strong nuclear point source would make bar detection more difficult.

Our result agrees with \citet{regan99}, \citet{martini99}, and
\citet{martini01}, who found
a low fraction of nuclear bars in Seyfert galaxies.
However, it contrasts with that of \citet{laine02} who find an ``excess of bars
among Seyfert galaxies at practically all length scales''.
Much of the difference may lie in our slightly more conservative
definition of bars; they use a maximum of 20$^\circ$ for the PA variation
in a bar, while we use $>10^\circ$ to define isophotal twists.
A detailed comparison of our bar classes with those in \citet{laine02}
supports this explanation.
Of the 34 Seyferts in common with them, our bar classifications
agree in 24 objects\footnote{We both find nuclear bars or none, or
we find no bars, and they find only large-scale bars.}.
Of the ten remaining objects, 7 of their nuclear-scale bars are defined
here as twists. 
Also, \citet{laine02} combine the high-resolution NICMOS images with large-scale
NIR and optical images, while
with our data, we are really only looking at nuclear bars.
Indeed, virtually all of the single bars, and a large fraction of the
secondary bars detected by \citet{laine02} would not be detected in our
images because of their small field-of-view.

\subsection{Boxy and Disky Isophotes \label{sec:boxydisky}}

The frequency of boxy isophotes (A4\,$<$\,0)
is not significantly different in any of the samples.
Boxiness in the inner kpc is rare, 
being found in $\sim$10\% of all galaxies (see Table \ref{tab:features}). 
 
Disky profiles (A4\,$>$\,0) are slightly more frequent than boxy ones.
Significantly disky isophotes are found in $\sim$20\% of all galaxies
except for \hii/starbursts (Table \ref{tab:features}).
Only 4$\pm$5\% of the \hii/starbursts have disky profiles, a significant
deficit relative to the other activity samples, but which 
rests on small-number statistics.  This absence of $\cos(4\theta)$
deviations in the starbursts' isophotes may be partly explained by
their unusually high frequency of $3\theta$ distortions, discussed next.

\subsection{3$\Theta$ Distortions and Dust Absorption \label{sec:3theta}}

Large $3\theta$ coefficients, A3/B3, occur most often in
\hii/starbursts (29\%).
They are least frequent in the non-active galaxies and LINERs 
($\sim10$\%), and intermediate in the Seyferts with $\sim16$\%.

The presence of strong deviations from elliptical isophotes, and in
particular large $3\theta$ coefficients,
are usually indicators of strong dust absorption \citep{peletier90}.
Such features are most frequent in the \hii\ galaxies, suggesting 
morphological disturbances traced by dust.
The Seyfert galaxies also show these disturbances, but only half as often
as \hii/starbursts, although large A3/B3 are more frequent in Seyferts
than in LINERs and non-active galaxies.

The A3/B3 residuals are the only fine structures that become significantly 
more likely in later galaxy morphologies. 
Indeed, our results show that, with the exception of the Sy2s,
by far the majority of A3/B3 residuals occur in spirals with
$T\geq3$.
To better assess whether our result depends on different
Hubble type distributions among samples, we have divided each
subsample into early ($T<3$) and late ($T\geq3$) as in $\S$\ref{sec:twists},
and performed again the statistical comparison.
We find that
among the late-type samples only, there is no significant difference
in A3/B3 residuals; any activity class of late Hubble type is equally likely
to show $3\theta$ deviations to smooth ellipses.
Among the early types however, in addition to the excess of A3/B3 residuals
in \hii\ galaxies (25\%), 20\% of the Sy2s but none of the Sy1s and only
4\% of the non-active galaxies
show $3\theta$ deviations; the high fraction in Sy2s
is a significant difference at 2.4$\sigma$. 
Hence, {\it in starbursts and Sy2s,
the excess of the A3/B3 features emerges
among the early Hubble types, where such morphologies are usually more rare.}

MGT also found dust absorption more often in Sy2s than
in Sy1s. There is some correlation between our finding of $3\theta$
residuals and their reporting dust, but it is far from perfect.
One reason for this may be because the shorter wavelength of the
{\it F606W} filter made the WFPC2 images analyzed by MGT more sensitive 
to dust lanes.
Another reason for the difference may be that dust lanes are not
necessarily distributed in ``banana-'' or ``heart-shaped'' isophotes;
our diagnostic would miss these.

\subsection{Unsharp-Mask Residuals \label{sec:usmresiduals}}

Half the \hii/starbursts show negative USM residuals, followed by
the type 2 Seyferts with 15\%,
11\% of the Sy1s, and 9\% of the non-active galaxies and LINERs.
The excess of negative USM residuals in the \hii\ galaxies is
highly significant.
A similar frequency (50\%) of positive USM residuals is found for 
the \hii/starbursts, followed by the Sy1s with 26\%, 17\% of the
non-active galaxies, 11\% of the Sy2s, and no LINERs.
Again the excess of positive USM residuals in the \hii\ galaxies
is significant.
The LINERs have the lowest frequency of features in the USM images--they
have the smoothest isophotes.

The statistics for the negative USM residuals are similar to the
$3\theta$ ones (see Fig. \ref{fig:usm}), lending support to the
idea that both diagnostics are revealing irregular dust morphology.
We repeated the statistical analysis for USMs by dividing each subsample
into early and late types as in the previous section.
Like the A3/B3 residuals,
the negative USM ones are more than 3 times as common
in late-type spirals as in early-type ones
independently of activity type, except for
the \hii/starbursts where they are equally as common. 
Among the early-type spirals 
the only three classes that show negative USM residuals 
are \hii's (67\%), Sy2s (9\%), and
non-active galaxies (4\%); no early-type LINERs nor
Sy1s have these features.
In terms of these diagnostics,
the \hii/starbursts are the galaxies most affected by dust.
Sy2s are the next most affected class, particularly among the
early Hubble types.

We checked that the excess of positive USM residuals in Sy1s
is not due to strong point-source contamination.
Visual inspection shows that the positive residuals are
more extended and irregular than a strong nuclear PSF.
Also most of the US mask structures are not aligned with the 
diffraction spikes of the central point source (see e.g., Fig. \ref{fig:imgsy1})
We therefore conclude that the positive USM residuals are real,
and not an artifact of the strong nuclear unresolved sources
generally seen in type 1 Seyferts.
It is noteworthy that a similar excess of positive USM residuals
is found in \hii/starburst galaxies; both Sy1s and starbursts
have significantly greater fractions of positive USM residuals
than any other class.
However, we hesitate to ascribe them to the same cause (localized bursts of
star formation?) because of the other differences
in morphology between the two classes.

\section{Non-axisymmetric Structure and AGN Fueling \label{sec:fueling}} 

The most robust result which emerges from our analysis is the
excess of isophotal twists in type 2 Seyferts.
Such features must be implicated in the fueling of BHs in Sy2s,
but not in Sy1s.
Isophotal twists in spiral galaxies can have several possible
causes. 
Projection effects on triaxial ellipsoids (bars, bulges) can cause the 
isophotes to appear twisted when viewed from an oblique angle
\citep{wozniak95,friedlietal,jungwiert}.
Independently of projection angle,
triaxial bulges embedded within a primary bar 
or nested misaligned secondary bars within primary bars can also 
result in twisted isophotes \citep{friedlimartinet,shaw95,elmegreen96}.
The presence of gas accumulated within orbital resonances in a barred
galaxy could also give rise to twists, because of the 
misalignment of the central stellar component with respect to the
primary bar \citep{shaw93,combes94,friedlietal,jungwiert}.
The models by \citet{knapen95}, \citet{heller96}, and \citet{regan03}
show how this could happen,
although perhaps not all Sy2s have a sufficiently high gas fraction
for nuclear gaseous disks to be a universally viable explanation.
On the other hand, the isophotal twists could be related to
the deficiency of thin stellar bars in Seyferts found 
by \citet{shlosman00}; bars tend to be weaker (thinner) in the
presence of a cold and clumpy gas component which could 
be causing the twists.

Our analysis is generally not able to distinguish among these alternatives,
although we argue that projection effects at these
spatial scales are difficult to correct for.
While some authors have deprojected ground-based images (e.g.,
\citealt{friedlietal,jungwiert,laine02}), we have not done so
because there is no {\it a priori} reason to suppose that
the circumnuclear structures probed by our images are coplanar with
the outer disk.

Star formation patterns, spiral arms, 
or absorption by dust could produce distorted central isophotes,
although this is much less probable in the NIR than in the optical.
However, a substantial number of the twisted isophotes in
Sy2s may be due to genuine nuclear triaxial structures.
Some of them are clearly associated with dust features, and
morphological disturbances signified by the
$3\theta$ coefficient (e.g., UGC\,2456=Mrk\,1066) or dust lanes
(e.g., NGC\,3079), but
others are found in otherwise unremarkable morphologies,
even in the USM image (e.g., NGC\,3982).
All of the structures that could give rise to isophotal twists
in our images
$-$triaxial bulges, nested misaligned bars, nuclear gas disks$-$
would disrupt kinematics at small spatial scales.
Nuclear disks and bars are associated with inward gas flow, but 
the connection between triaxiality and inflow is not so clear. 
Either way {\it twists appear to be a signature of Seyfert activity, but only in 
type 2s}.
Possible reasons why isophotal twists are not found as frequently in Sy1s 
are discussed below.

What we identify as isophotal twists could also be a manifestation of 
the nested-bar scenario of \citet{shlosman89} and \citet{laine02}.
Nevertheless, since only a fraction of the  twists in our sample are 
found in barred galaxies, our observations
may not be entirely consistent with it.

We find significant differences in the dust content of Sy1s and Sy2s
{\it among the early Hubble types}, in agreement with \citet{storchi}.
Among late types, 
there is no difference in A3 or negative USM residuals,
where they are more frequent in all activity classes.
However, among early types the differences between Sy2s and Sy1s are 
significant. 
Seyferts are also found primarily in early morphological types \citep{moles},
which suggests that Hubble type may be related to nuclear activity.
However, type alone does not seem to be enough.
Seyfert activity, at least in Sy2s,
seems to be distinguished also by a greater morphological irregularity, 
as shown by the excess of twists and A3s. 

\section{Testing Unification Schemes: 
Comparison of Seyfert 1 and 2 Host Galaxies }

Unified Schemes assert that Seyfert 1 and 2 nuclei are intrinsically the same.
Their apparent differences are due to additional dust absorption of much of
the Seyfert 1 emission (UV to soft-Xray continuum and broad emission lines).
In the torus model, this absorption occurs very close to the central engine
and is co-aligned with its axis.  Since this axis is in general uncorrelated 
with the major axis of the host galaxy \citep{keel,schmitt01},
no systematic differences are expected between Seyfert 1 and Seyfert 2 host
galaxy properties.

In apparent disagreement with the torus model, 
we find some significant differences between our samples of 
Seyfert 1 and 2 host galaxies.

\begin{itemize}
\item
Isophotal twists are twice as common in Sy2s as in Sy1s
(63$\pm$7\% vs. 30$\pm$9\%). 
This significant difference (see $\S$\ref{sec:twists}) is not readily
explained by the Unified Scheme.
\item
Bars are present in 15\% of our Sy2 galaxies, but in 2\% 
of the Sy1s.
Even if we loosened our definition of a ``bar", so that it would 
include the cases we call ``isophotal twists", the Sy 1 vs. 2
difference would still remain: Sy2s have a significantly higher
fraction of bars or twists than do Sy1s.
\item
Among the early-type Seyferts ($T<3$), 20\% of the Sy2s
have $3\theta$ residuals, but no Sy1s, a formally significant difference
(see $\S$\ref{sec:3theta}).
Nevertheless,
our other dust indicator, strong negative residuals in the USMs, 
confirms only weakly 
the suggestion of excess dust in the centers of Sy2s. 
Our result is weaker than that of
MGT who claimed that the centers of Sy2 galaxies had systematically more,
or more widely distributed dust absorption than those of Sy1 galaxies.
The reason may be due to the 
lack of sensitivity to dust of our infrared images,
although sample effects may also be important; we find
no differences among late Hubble types but significant ones among early types.
\end{itemize}


If the Seyfert 1 and 2 galaxies differ in more ways than just the
orientation of a central torus, then perhaps they
represent nuclear activity from black hole accretion in
different evolutionary stages, which we will now explore.

\subsection{Evolutionary Scenario? \label{sec:evolution}}

Our sample of ``normal" (non-active) galaxies defines a morphological
baseline against which various samples of active galaxies can be compared.
We find that the central isophotes of normal galaxies are usually well described
by ellipses at a relatively constant (to within 10$^\circ$) position angle.
Thus they show very few significant $3\theta$ or $4\theta$ deviations, or 
other large excesses or deficiencies of light that would appear in unsharp masking.
The other extreme of our non-AGN baseline is defined by the \hii/starburst
galaxies which contain relatively large amounts of interstellar matter and young stars.
These differ mainly from the normal galaxies in having strong positive and
negative light excesses in their USMs, and strong $3\theta$ deviations
from elliptical isophotes.

Among our AGN (``active'') galaxy sub-samples, the LINERs show the 
least non-axisymmetric structure, and the closest morphological
similarity to the normal galaxies. 
They have the smallest fraction of significant A3 residuals, and  
the USMs show that they are in fact even more featureless
than the normal galaxies.
That the LINER sample is the closest most well-resolved one
only strengthens this result.
This would be consistent with the view that LINERs are early-type
galaxies with little indications of any recent disturbance.  
The Seyferts lie
between the \hii/starbursts and the LINERs. But we are unable to identify any
particular morphological peculiarities that equally distinguish {\it both}
Seyfert 1s and Seyfert 2s from the other kinds of less active galaxies.
Thus one standard morphological ``explanation" for {\it all} Seyfert activity
may not exist. 

We speculate that
these patterns of small-scale non-axisymmetric structure are the footprints
of an evolutionary scenario, 
which we hypothesize starts with a \hii/starburst, and ends with 
(or returns to) a normal galaxy.  
\hii/starburst galaxies are morphologically ``younger",  
viewed soonest after 
the onset of a dynamical instability, either intrinsic or extrinsic 
induced by a merger, interaction, or accretion event. 
Large A3/B3 and USM residuals are the morphological signatures of
the cataclysmic perturbation(s) which triggered the starburst.  
Nuclear bars on the 100s-pc scales studied here do not seem to be
directly implicated for the starburst, since the bar fraction in the
\hii\ galaxies is not excessive, unlike their fraction of large-scale bars 
\citep{huntmalkan99}.
 
Our results would place Sy2s earlier or ``younger"
than Sy1s in the evolutionary sequence, but later or ``older'' morphologically
than a \hii/starburst.
Of the two Seyfert types, Sy2s appear to be more morphologically disturbed 
in their central regions than Sy1s.
More than 60\% of the Sy2s show isophotal twists, and
their A3/B3 fraction in the early-type subset (see $\S$\ref{sec:3theta})
is second only to the \hii\ galaxies. 
Sy2s are also intermediate between starbursts and Sy1s in other
features such as disky isophotes and nuclear bars.
The only exception to this 
generalization is the higher incidence of positive USM features in 
Sy1s and starbursts, for which we have no explanation. 

Our hypothetical placement of Seyferts as intermediate evolutionary stages between
``younger" starburst galaxies and ``older" non-active galaxies and LINERs is consistent 
with our previous results from morphology on larger scales.
\hii/starbursts in the 12\,$\mu$m sample were found to have an excess of large-scale bars, 
and Seyferts unusually high rates of outer rings \citep{huntmalkan99}. 
Either of these features 
could be produced by some instability or interaction event, but 
outer rings cannot even form before $10^9$ yr and require a bar
to do so \citep{bc}.  This would suggest
that Seyfert activity is prompted by a disturbance, but with a significant
time delay relative to the relatively rapid burst of star formation that preceded it.
The structures that could be responsible for twists
(nuclear disks, nested or misaligned bars) have 
evolution times of a few $\times10^8$ \,yr, but the process
requires a bar to have already formed (at least in the simulations, see 
\citealt{shaw93,knapen95,heller96,friedlietal}).
Hence, isophotal twists are apparently younger than outer rings, but
older than the relatively prompt results of a galaxy interaction, such as 
a violent star-formation episode.
Twists may be a subsequent phase of what originally began as a bar-induced
starburst.

Taken together, these results could be a confirmation of the
\hii-Sy2 evolutionary scenario proposed by \citep{storchi,kauffmann03}, 
and the \hii-Sy2-Sy1 scenario proposed by us and other groups 
\citep{huntmalkan99,krongold,levenson}.
Relatively young ($\simlt$1\,Gyr) stellar populations are found in 
more than half of type 2 Seyferts 
\citep{schmitt99,gonzales,cidfernandes,raimann}. 
High-luminosity broadlined AGN in general host 
similarly young populations, and there is evidence for bursts of star formation
in AGNs which occurred up to a few Gyr ago \citep{kauffmann03}. 
These timescales agree roughly with what we have deduced from the 
non-axisymmetric morphology.
First, a dynamical instability turns a galaxy into a starburst 
with disturbed and dusty morphology, on a timescale corresponding to the 
large-scale bar formation time of $\sim10^8$\,yr.
Given a sufficient gas supply, a starburst could evolve toward
a Sy2, after the time necessary (few times $10^8$\,yr) to set up 
sufficiently efficient gas inflow as manifested by nuclear disks, 
misaligned bars, or triaxial structures, to which we attribute the observed twists.
After another Gyr or two, a Sy2 could ``settle down'' to a Sy1
with an outer ring which is a signature of the previous inflow and consequent
outflow of material and angular momentum.
Such an evolutionary trend could be episodic, depending on the environment
and the disk kinematics, since a new instability or external
perturbation could start the process all over again. 

LINERS are the most morphologically ``settled down", 
viewed perhaps much after the event which triggered a starburst,
and, in our picture, the onset of nuclear activity.
This would imply that LINERs are either unrelated to Seyferts, or are
``exhausted" Seyferts at the end of their fuel supply.
The latter scenario seems more plausible in the light
of the significant excess of {\it inner} rings in LINERs \citep{huntmalkan99}. 
In inner or nuclear rings, gas tends to 
pile up in the resonances rather than funneling inward
to the nucleus \citep{regan03}.
Thus inner rings slow down or halt completely the gas
supply available for feeding an accreting BH (e.g., \citealt{combes04}).
LINERs, with their high fraction of inner rings, could be
``starving AGNs''.

\section{Conclusions}

NICMOS imaging of the centers of large numbers of normal and
active galaxies has revealed some systematic morphological differences.
The normal galaxies and LINERs tend to have the most regular images, while
\hii/starbursts are the most disturbed. 
The Seyfert galaxy morphologies tend to be intermediate between
these two extremes.

Sy2s appear to be more structurally relaxed than \hii/starbursts,
but are  more disturbed than Sy1s and LINERs.  In terms of
circumnuclear peculiarities, Sy2s appear to be intermediate between
\hii/starbursts and Sy1s;
they show substantially more inner isophotal twists than any other class,
and, in the early-type subset, 
are between the \hii\ galaxies and the Sy1s in terms of $3\theta$ and USM
residuals.

If we hypothesize that the non-axisymmetric structure
in the central part of the galaxy can
influence the active nucleus by enhancing its gas fueling rate,
then our morphological data can fit into
the evolutionary scenario advanced by us \citep{huntmalkan99}
and other groups \citep{krongold,levenson}.
The first result of a dynamical instability,
perhaps caused by an interaction/close encounter, is to transform a
``normal" galaxy into a ``starburst"; then for a few hundred million years
its spectrum may be characterized as an ``\hii\ region".
Later, as the bar evolves, perhaps forming
nuclear disks or nested bars, the galaxy is more likely to appear 
as a Seyfert 2 because of increased gas inflow to the nuclear BH.
Finally, after another billion years or so, when the central structure
has evolved and ``relaxed'' into a greater degree of axisymmetry,
but still able to feed the BH, the galaxy would appear as a Seyfert 1.
LINERs, because of their smooth appearance, 
could be ``starved'' AGNs, having
exhausted the available fuel supply, at least temporarily.

\acknowledgments
We thank Wayne Webb for help with the preliminary recalibration of some of the images.
We are grateful to Reynier Peletier for carefully reading and commenting
on the manuscript, and to an anonymous referee whose insight and critiques
greatly improved the paper.
This work has relied heavily on the NASA Extragalactic Database.
Support was received from NASA/STScI grant GO-7328.

\appendix
\section{Distributions of Parameters for Matched Samples \label{sec:matched}}

We have analyzed the full sample of 250 galaxies in terms
of distance, physical spatial scale, luminosity, Hubble type,
and inclination in order to minimize possible selection biases
among the activity samples.
The properties of the galaxies which we have eliminated in order to construct
the matched samples described below are given
in Table \ref{tab:appendix}.

\subsection{Distance}

To ensure sufficiently high spatial resolution,
we eliminated from further consideration all
NIC-3 galaxies with recession velocities $>$\,5000\,km\,s$^{-1}$.
Pixel size in parsecs (\ppc) was then obtained from the camera pixel scales,
with distances determined 
by assuming a Hubble constant of $H_0=72$\,km\,s$^{-1}$\,Mpc$^{-1}$
\citep{shoko}, and a Virgocentric infall model with
an infall velocity of -300\,km\,s$^{-1}$ \citep{geller}.
The distances of four nearby galaxies 
(NGC\,404, NGC\,4395, NGC\,4569, NGC\,5055) were taken
from the Nearby Galaxies Catalogue \citep{tully},
and other galaxies with negative velocities were assigned a distance 
of 5\,Mpc.

All objects with distances $<$ 10\,Mpc were eliminated a priori.
Moreover,
to match the distance medians of the subsamples to within $\sim$20\,Mpc, we had to 
eliminate several distant Seyfert galaxies. 
The resulting set of samples (to which all the constraints described
in the following sections also apply) contains 165 galaxies, and
we denote it as MS (Matched Samples).
The worst mismatch is that between the LINERs and Sy1s;
the Sy1 sample is roughly $\Delta D\,\sim$\,27\,Mpc (55\%)
more distant than the LINERs. 
The distance medians of the remaining samples relative to the
Seyferts are within $\sim$\,20\,Mpc of one another. 
Both Seyfert samples are, at median distances of 49 (Sy1s) and 37 (Sy2s) Mpc, 
substantially closer than either the 12\,$\mu$m
sample \citep{huntmalkan99}, or the CfA Seyferts \citep{huchraburg}.

We also created a second more rigorous set of samples (DMS, Distance Matched Samples)
consisting of 152 galaxies,
in which all objects with distances $\geq$\,80\,Mpc were thrown out.
4 Sy2s and 9 Sy1s were eliminated by this constraint.
The median distances of the DMS are much more closely matched,
with 38\,Mpc for the Sy1s and 36\,Mpc for the Sy2s
(see Table \ref{tab:medians}).

\subsection{Parsec-per-pixel Scale \label{sec:ppc}}

It is even more important to match our images to comparable resolution
scales.
Because the NIC-3 pixels are so large compared
to the other two cameras (0.20\arcsec\ vs. 0.075\arcsec\ and 0.0475\arcsec), 
we truncated the samples so that no 
activity sample contained galaxies with pc/pixel\,$\geq$30.
This cutoff, while arbitrary, defines roughly the physical scales of 
the morphological features we are interested in, since 3 pixels would
correspond to $\sim$\,100\,pc in the worst case.
The majority of the galaxies excluded with this criterion were non-active
ones, given that many of them were imaged with NIC-3.
Constraining the parsec-to-pixel scale eliminated 28 galaxies, 23 of which
are non-active.
Both the MS and the DMS mentioned above and the third sample described below
include this constraint.

\begin{figure*}[t]
\centering
{\includegraphics[width=0.6\linewidth, bb=30 144 440 715]{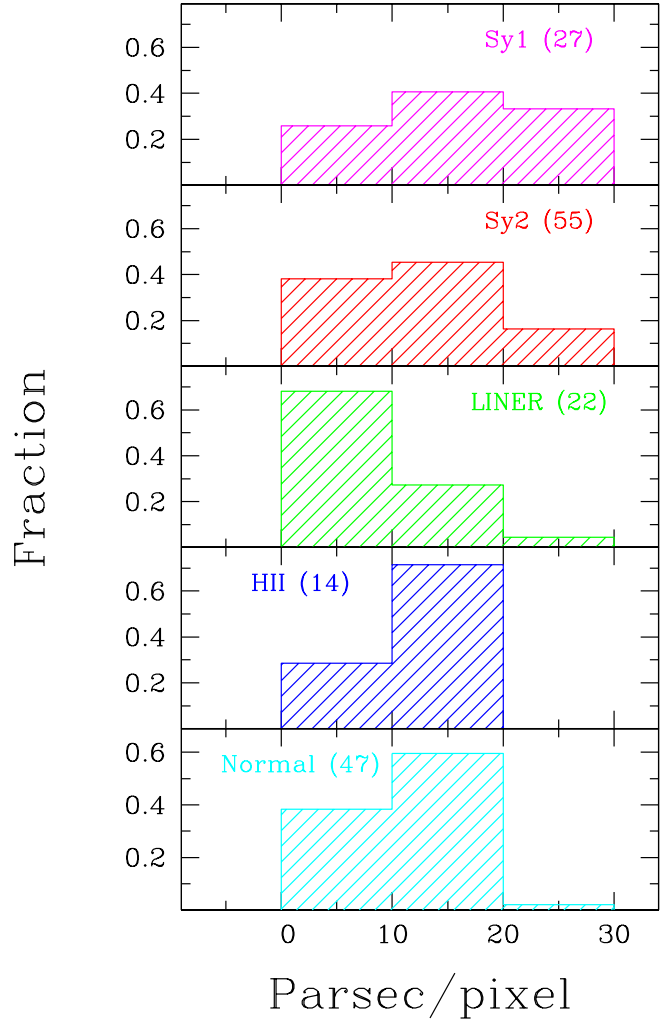}}
\caption{Distribution of spatial resolution (parsec/pixel)
for each activity class.
The binning of 10pc/pixel is deceptive, as the maximum
resolution is $\sim$25pc/pixel (see text).
\label{fig:p2pc}}
\end{figure*}

The median physical pixel scales range from 7.5 (LINERs) to
15.2 (Sy1s) pc/pixel.
In the DMS, the median Sy1 scale is 13.2 pc/pixel.
The remaining samples have similar medians with
10.4, 12.6, and 12.9 pc/pixel, for the non-active, \hii, and Sy2 samples,
respectively.
The maximum pixel scale is 25.9 pc/pixel for the Sy2 sample, closely
followed by the Sy1s with 24.9 and the non-active sample with 24.5 pc/pixel.
Figure \ref{fig:p2pc} shows the relative parsec-to-pixel spatial scales
for the MS.

\placefigure{fig:p2pc}

\subsection{Blue Luminosity }

It was necessary to remove the lower luminosity galaxies from the normal
and \hii\ subsamples, because of the presence of (25) dwarf galaxies.
The final minimum in blue absolute magnitude $M_B$ is $-18.4$, which
resulted in a worst discrepancy of $\Delta M\,\sim$\,1\,mag between the normal
sample and the more luminous Seyfert ones.
The high-luminosity Seyfert galaxies were eliminated from the sample with
the distance and \ppc\ criteria described previously.
The low- and high-luminosity ends of all samples turn out to be similar 
($\sim\,-18.5$ and $\sim\,-21.5$) with the 
most and least luminous galaxies both being LINERs (NGC\,1961: $-22.0$ 
and NGC\,5879: $-18.4$).
These ranges are virtually identical to the CfA Seyfert sample
\citep{huchraburg}, and we conclude that our samples
should be a fair representation of Seyfert galaxies in terms of luminosity.

\subsection{Hubble Type}

The earliest types present have T=$-4$, one exemplar of which is present 
in the LINER (NGC\,3379\,=\,M\,105) and Seyfert (NGC\,1275, NGC\,4278) samples.
The latest types in the samples are one exemplar of T=9 
(normal), and one of T=7 (\hii).
The remaining galaxies range from T=$-2$ to T=6.
Two Seyfert galaxies were classified as ``S'' (spiral) in RC3, and we
assigned T=5 to these (see the LEDA database).
The sample medians for Hubble type are Sab (LINER and Sy2),
Sb (normal and Sy1), and Sbc (\hii).
These are very similar to the type distributions found in the 
12\,$\mu$m sample \citep{huntmalkan99}, except for the later types here 
of Sy1s (12\,$\mu$m Sy1s have median Sa) and earlier types here of LINERs
(12\,$\mu$m LINERs have median Sbc).
The normal galaxies have the same median Hubble type as the Sy1 sample,
and lie between the remaining active subsamples; thus
they should be a ``fair'' comparison. 

\subsection{Galaxy Inclination}

Because the inclination of a galaxy may affect our ability to distinguish
morphology, we also checked for the apparent axial ratio distributions.
The normal galaxies have a median inclination of 63$^\circ$, while the remaining
samples $\sim\,50^\circ$.
In particular the Sy2 sample has a median axial ratio of 0.73 (43$^\circ$),
which may reflect the paucity of optically selected edge-on Seyferts \citep{keel}.

In order to ensure that we are not missing peculiar morphology because of   
excesses in galaxy inclination,
we created a third sample set of 139 galaxies,
denoted as IMS (Inclination Matched Samples). 
Here all galaxies with inclination $>$\,70$^\circ$ were eliminated.
This operation had the greatest impact on the non-active sample in which
15 galaxies were excluded.

\subsection{Bar Types \label{sec:bartypes}}

Finally,
we investigated the bar classifications based on the RC3 designations.
With the exceptions of the \hii\ sample which shows an excess of bars similar
to the Markarian and 12\,$\mu$m starbursts \citep{huntmalkan99}
(84\% strongly-SB or weakly-SAB barred) and the LINER sample with a deficit of bars
(52\% unbarred), the remaining three samples (non-active and Seyferts) show
``normal'' bar properties \citep{moles,ho97}
with $\sim 60-70$\% of the galaxies being either SB or SAB.


\begin{deluxetable}{lrrlrrrrrlllll} 
\rotate
\tablecolumns{14} 
\tableheadfrac{0.05} 
\tabletypesize{\scriptsize}
\tablewidth{0pt}
\tablenum{4}
\tablecaption{Properties of Rejected Galaxies\label{tab:appendix}}
\tablehead{ 
\multicolumn{1}{c}{Name} & 
\multicolumn{1}{c}{$z$} & \multicolumn{1}{c}{Dist.} &
\multicolumn{1}{c}{RC3 Type} &  \multicolumn{1}{c}{a} &
\multicolumn{1}{c}{b} &  \multicolumn{1}{c}{Mag.} &  \multicolumn{1}{c}{Abs. Mag.} &
\multicolumn{1}{c}{Pc/pixel} & \multicolumn{1}{c}{Bar} & 
\multicolumn{1}{c}{Twist} & \multicolumn{1}{c}{3$\theta$} & 
\multicolumn{1}{c}{cos(4$\theta$)} & \multicolumn{1}{c}{USM} \\
\colhead{(1)} &  \colhead{(2)} & 
\colhead{(3)} & \colhead{(4)} &
\colhead{(5)} & \colhead{(6)} &
\colhead{(7)} & \colhead{(8)} & 
\colhead{(9)} & \colhead{(10)} & 
\colhead{(11)} & \colhead{(12)} & 
\colhead{(13)} & \colhead{(14)} } 
\startdata
\cutinhead{Non-active}
ESO205-G007          & 0.007 &  26.8 & SAB(rs)bc       &  0.9 &  0.8 &  15.2 & $-16.9$  &  9.7 &    & Y? &    & D  &    \\
ESO240-G012          & 0.006 &  22.3 & S?              &  1.3 &  0.5 &  14.3 & $-17.4$  &  8.1 &    &    &    &    &    \\
ESO548-G029          & 0.004 &  13.8 & SB?             &  1.1 &  0.8 &  14.3 & $-16.4$  &  5.0 &    &    &    &    &    \\
ESO549-G018          & 0.005 &  19.2 & SAB(rs)c        &  2.6 &  1.5 &  13.7 & $-17.7$  &  7.0 &    &    &    &    &    \\
ESO572-G022          & 0.006 &  30.4 & Sb              &  1.3 &  0.4 &  14.8 & $-17.6$  & 11.1 &    &    &    &    & $+$ \\
IC0749               & 0.003 &  14.6 & SAB(rs)cd       &  2.3 &  1.9 &  12.9 & $-17.9$  & 14.2 & Y  & Y? &    &    &    \\
IC0750               & 0.002 &  13.4 & Sab:sp          &  2.6 &  1.2 &  12.9 & $-17.7$  & 13.0 &    & Y  &    &    & $+$ \\
MESSIER074           & 0.002 &   5.4 & SA(s)c          & 10.5 &  9.5 &   9.9 & $-18.8$  &  5.2 &    & Y  &    &    &    \\
NGC0151              & 0.012 &  47.7 & SB(r)bc         &  3.7 &  1.7 &  12.3 & $-21.1$  & 46.3 & Y  & Y  &    &    &    \\
NGC0214              & 0.015 &  59.4 & SAB(r)c         &  1.9 &  1.4 &  13.0 & $-20.9$  & 57.6 &    & Y  & Y  &    &    \\
NGC0491              & 0.013 &  49.8 & SB(rs)b:        &  1.4 &  1.0 &  13.2 & $-20.3$  & 48.3 & Y  & Y  & Y  &    &    \\
NGC1345              & 0.005 &  18.1 & SB(s)cpec:      &  1.5 &  1.1 &  14.3 & $-17.0$  &  6.6 &    &    &    &    & $+$ \\
NGC1483              & 0.004 &  13.5 & SB(s)c          &  1.6 &  1.3 &  13.1 & $-17.5$  &  4.9 &    &    & Y? &    &    \\
NGC1688              & 0.004 &  15.5 & SB(rs)dm        &  2.4 &  1.9 &  12.6 & $-18.4$  &  5.6 &    &    &    &    & $+$ \\
NGC2082              & 0.004 &  14.2 & SAB(rs+)c       &  1.8 &  1.7 &  12.6 & $-18.2$  &  5.2 &    & Y? &    &    &    \\
NGC2104              & 0.004 &  15.0 & SAB(s)cd:       &  2.0 &  0.9 &  13.2 & $-17.7$  &  5.5 &    &    &    &    &    \\
NGC2314              & 0.013 &  54.1 & E3              &  1.7 &  1.4 &  13.2 & $-20.5$  & 52.5 &    & Y? &    &    &    \\
NGC2344              & 0.003 &  14.7 & SA(rs)c:        &  1.7 &  1.7 &  12.8 & $-18.0$  &  5.3 &    &    &    &    &    \\
NGC2642              & 0.014 &  62.6 & SB(r)bc         &  2.0 &  1.9 &  13.3 & $-20.7$  & 60.7 & Y? & Y  &    & D? &    \\
NGC2672              & 0.014 &  62.9 & E1-2            &  3.0 &  2.8 &  12.7 & $-21.3$  & 61.0 &    & Y  &    &    &    \\
NGC2749              & 0.014 &  61.2 & E3              &  1.7 &  1.4 &  12.7 & $-21.2$  & 59.3 &    &    &    & B? &    \\
NGC2758              & 0.007 &  29.6 & (R')SBbcpec?    &  1.9 &  0.5 &  14.0 & $-18.4$  & 10.8 &    &    &    &    & $-$ \\
NGC2942              & 0.015 &  64.5 & SA(s)c:         &  2.2 &  1.8 &  13.2 & $-20.9$  & 62.5 &    & Y  &    &    &    \\
NGC2998              & 0.016 &  69.4 & SAB(rs)c        &  2.9 &  1.3 &  13.1 & $-21.1$  & 67.3 & Y  & Y  &    &    &    \\
NGC3271              & 0.012 &  54.5 & SAB(s)0$^0$       &  3.1 &  1.8 &  12.9 & $-20.8$  & 52.8 &    & Y  &    &    &    \\
NGC3275              & 0.011 &  47.1 & SB(r)a          &  2.8 &  2.1 &  11.8 & $-21.6$  & 45.7 & Y  & Y  &    &    &    \\
NGC3544              & 0.012 &  53.8 & (R)SAB:(rs)a    &  3.0 &  1.0 &  13.0 & $-20.6$  & 52.2 &    &    &    &    &    \\
NGC3769              & 0.002 &  13.6 & SB(r)b:         &  3.1 &  1.0 &  12.6 & $-18.1$  & 13.2 & Y? & Y  &    &    & $+$ \\
NGC3782              & 0.002 &  13.8 & SAB(s)cd:       &  1.7 &  1.1 &  13.1 & $-17.6$  & 13.4 &    &    &    &    & $+$ \\
NGC3928              & 0.003 &  17.1 & SA(s)b?         &  1.5 &  1.5 &  13.2 & $-18.0$  &  6.2 &    &    & Y? &    &    \\
NGC4085              & 0.003 &  13.8 & SAB(s)c:?       &  2.8 &  0.8 &  12.9 & $-17.8$  & 13.4 &    & Y  &    &    &    \\
NGC4373              & 0.011 &  50.0 & SAB(rs)0-:      &  3.4 &  2.5 &  11.9 & $-21.6$  & 48.5 &    &    &    &    &    \\
NGC4701              & 0.002 &  14.4 & SA(s)cd         &  2.8 &  2.1 &  12.8 & $-18.0$  & 14.0 &    & Y  &    &    & $-$ \\
NGC4786              & 0.016 &  68.7 & E+pec           &  1.6 &  1.3 &  12.7 & $-21.5$  & 66.6 &    &    &    &    &    \\
NGC5444              & 0.013 &  59.2 & E+:             &  2.4 &  2.1 &  12.8 & $-21.1$  & 57.4 &    &    &    &    &    \\
NGC5605              & 0.011 &  50.6 & (R')SAB(rs)cpec: &  1.6 &  1.3 &  13.2 & $-20.3$  & 49.1 &    & Y  &    &    &    \\
NGC5641              & 0.014 &  64.0 & (R')SAB(r)ab    &  2.5 &  1.3 &  13.1 & $-20.9$  & 62.1 &    & Y? & Y  & D  &    \\
NGC5908              & 0.011 &  48.4 & SA(s)b:sp       &  3.2 &  1.2 &  12.8 & $-20.6$  & 46.9 &    &    &    & B  &    \\
NGC6699              & 0.011 &  46.5 & SAB(s)bc        &  1.5 &  1.5 &  12.6 & $-20.7$  & 45.1 & Y  & Y  &    &    &    \\
NGC6754              & 0.011 &  44.1 & SAB(rs)bc       &  1.9 &  0.9 &  12.9 & $-20.3$  & 42.8 & Y? & Y  &    & D  &    \\
NGC6808              & 0.012 &  47.0 & SA(r)abpec:     &  1.5 &  0.8 &  12.5 & $-20.9$  & 45.6 &    & Y? &    &    &    \\
NGC6876              & 0.013 &  54.3 & SB0$^-$           &  2.8 &  2.2 &  12.1 & $-21.6$  & 52.7 &    &    &    &    &    \\
NGC7188              & 0.006 &  20.9 & (R'$_2$)SB(s)bc &  1.6 &  0.7 &  13.8 & $-17.8$  &  7.6 &    &    &    &    &    \\
NGC7259              & 0.006 &  20.2 & Sb              &  1.1 &  0.9 &  13.9 & $-17.6$  &  7.3 &    &    &    &    &    \\
NGC7309              & 0.013 &  51.7 & SAB(rs)c        &  1.9 &  1.8 &  13.0 & $-20.6$  & 50.1 & Y  & Y  &    &    &    \\
NGC7457              & 0.003 &   8.2 & SA(rs)0-?       &  4.3 &  2.3 &  12.1 & $-17.5$  &  3.0 &    &    &    &    &    \\
NGC7513              & 0.005 &  17.7 & (R')SB(s)bpec   &  3.2 &  2.1 &  13.1 & $-18.1$  &  6.4 &    & Y? &    &    &    \\
NGC7690              & 0.005 &  17.9 & Sb              &  2.2 &  0.9 &  13.0 & $-18.3$  &  6.5 &    &    &    &    &    \\
UGCA196              & 0.003 &  16.1 & (R')SA(s)b      &  3.2 &  1.3 &  13.3 & $-17.7$  &  5.9 &    &    &    &    &    \\
\cutinhead{HII}
IC0745               & 0.004 &  20.2 & S0              &  0.7 &  0.6 &  14.2 & $-17.3$  &  7.3 &    & Y  & Y  &    &    \\
IC4870               & 0.003 &  11.1 & IBm?pec         &  1.6 &  0.9 &  13.9 & $-16.3$  &  4.0 &    &    &    &    &    \\
MRK0930              & 0.018 &  72.9 & Pair            &  0.0 &  0.0 &  17.0 & $-17.3$  & 26.5 &    &    &    &    & $+$ \\
NGC2989              & 0.014 &  60.4 & SAB(s)bc:       &  1.7 &  0.9 &  13.6 & $-20.3$  & 58.6 & Y  & Y  &    &    &    \\
NGC5653              & 0.012 &  53.1 & (R')SA(rs)b     &  1.7 &  1.3 &  12.9 & $-20.7$  & 51.5 &    & Y  &    &    &    \\
\cutinhead{LINER}
MESSIER063           & 0.002 &   7.2 & SA(rs)bc/       & 12.6 &  7.2 &   9.3 & $-20.0$  &  7.0 &    &    &    &    &    \\
NGC0404              & -0.000 &   2.4 & SA(s)0-:        &  3.5 &  3.5 &  11.2 & $-15.7$  &  0.9 &    &    &    &    &    \\
NGC7013              & 0.003 &   8.9 & SA(r)0/a        &  4.0 &  1.4 &  12.4 & $-17.3$  &  3.2 & Y? &    &    & D? &    \\
NGC7331              & 0.003 &   8.6 & SA(s)b          & 10.5 &  3.7 &  10.3 & $-19.4$  &  3.1 &    &    &    & B?D? &    \\
\cutinhead{Seyfert 2}
CGCG164-019          & 0.030 & 128.0 & Sa              &  0.4 &  0.3 &  15.3 & $-20.2$  & 26.7 & Y? & Y  & Y  & D? & $-$ \\
MESSIER061           & 0.005 &  26.1 & SAB(rs)bc       &  6.5 &  5.8 &  10.2 & $-21.9$  &  9.5 & Y  & Y  &    &    &    \\
MESSIER096           & 0.003 &  16.4 & SAB(rs)ab       &  7.6 &  5.2 &  10.1 & $-21.0$  &  6.0 &    & Y? &    & D? &    \\
MRK0078              & 0.037 & 156.2 & SB              &  0.4 &  0.2 &  15.0 & $-21.0$  & 56.8 &    &    &    & D  &    \\
MRK0477              & 0.038 & 160.2 & S               &  0.0 &  0.0 &  17.0 & $-19.0$  & 58.3 &    & Y  & Y  & D  &    \\
NGC3362              & 0.028 & 119.1 & SABc            &  1.4 &  1.1 &  13.5 & $-21.9$  & 24.8 &    &    &    &    &    \\
NGC4117              & 0.003 &  16.8 & S0$^0$         &  1.8 &  0.9 &  14.0 & $-17.1$  &  6.1 &    &    &    &    &    \\
NGC7319              & 0.023 &  91.0 & SB(s)bcpec      &  1.7 &  1.3 &  14.1 & $-20.7$  & 33.1 &    & Y  & Y  & D  &    \\
UGC06100             & 0.030 & 126.2 & Sa?             &  0.8 &  0.5 &  14.3 & $-21.2$  & 26.3 &    & Y  &    &    &    \\
UGC06527             & 0.028 & 119.1 & SA(s)0/apec:    &  0.4 &  0.3 &  17.0 & $-18.4$  & 43.3 &    & Y? & Y  & D  &    \\
UGC12348             & 0.025 & 102.0 & Sa              &  1.0 &  0.3 &  15.3 & $-19.7$  & 37.1 &    &    &    & B  &    \\
UM625                & 0.025 & 108.0 & S0              &  0.3 &  0.2 &  17.4 & $-17.8$  & 22.5 &    & Y  &    &    &    \\
\cutinhead{Seyfert 1}
ARP151               & 0.021 &  90.9 & S0pec           &  1.3 &  0.2 &  16.8 & $-18.0$  & 18.9 &    & Y? &    & D  &    \\
IC1854               & 0.031 & 126.1 & S0/a            &  0.5 &  0.4 &  14.9 & $-20.6$  & 26.3 &    & Y  &    &    &    \\
MESSIER104           & 0.003 &  18.3 & SA(s)a          &  8.7 &  3.5 &   9.0 & $-22.3$  &  6.7 & Y? &    &    & D? & $+$ \\
NGC0985              & 0.043 & 175.7 & SBbc?p(Ring)    &  1.0 &  0.9 &  13.8 & $-22.4$  & 63.9 &    & Y  & Y  &    &    \\
NGC4395              & 0.001 &   3.6 & SA(s)m:         & 13.2 & 11.0 &  10.6 & $-17.2$  &  1.3 &    &    &    &    &    \\
NGC5940              & 0.034 & 144.2 & SBab            &  0.8 &  0.8 &  14.3 & $-21.5$  & 30.1 &    & Y  &    &    &    \\
UGC09214             & 0.034 & 146.0 & SBa             &  0.9 &  0.6 &  14.5 & $-21.3$  & 30.4 & Y  & Y  &    &    & $+$ \\
UGC10120             & 0.031 & 133.0 & SB(r)b          &  1.1 &  1.1 &  14.6 & $-21.0$  & 27.7 &    &    &    &    &    \\
\enddata 
\end{deluxetable}

\clearpage

\begin{figure*}[t]
\centering
\caption{INSERT HERE f2.gif.
Representative data for the normal (non-active)
sample: NICMOS {\it F160W} images (left panel);
elliptically averaged surface brightness profiles (middle panel); 
unsharp masked images (see text) (right panel).
In the images (left), North is up and East to the left.
In the USMs (right), white indicates positive excesses, and black negative ones.
Profiles are more completely described in the caption of \ref{fig:comp}.
\label{fig:imgnrm}}
\end{figure*}

\begin{figure*}[t]
\centering
\caption{INSERT HERE f3.gif.
Representative data for the \hii/starburst sample.
The presentation is the same as in Fig. \ref{fig:imgnrm}.
\label{fig:imghii}}
\end{figure*}

\begin{figure*}[t]
\centering
\caption{INSERT HERE f4.gif.
Representative data for the LINER sample.
The presentation is the same as in Fig. \ref{fig:imgnrm}.
\label{fig:imglin}}
\end{figure*}

\begin{figure*}[t]
\centering
\caption{INSERT HERE f5.gif.
Representative data for the Sy2 sample.
The presentation is the same as in Fig. \ref{fig:imgnrm}.
\label{fig:imgsy2}}
\end{figure*}

\begin{figure*}[t]
\centering
\caption{INSERT HERE f6.gif.
Representative data for the Sy1 sample.
The presentation is the same as in Fig. \ref{fig:imgnrm}.
\label{fig:imgsy1}}
\end{figure*}

\begin{figure*}[t]
\centering
\caption{INSERT HERE f7.gif. Comparison of elliptically averaged surface
brightness profiles for different observers or different NICMOS
cameras.  The panels plot the radial run of surface brightness,
ellipticity $\epsilon$,
position angle $\theta$, $3\theta$ residuals (A3/B3), and
$4\theta$ residuals (A4/B4). 
In the upper panel, the solid line corresponds to the best-fit
Nuker function (not discussed in this paper), and the dotted
lines to the galaxy and nuclear components. 
In the lower panels (A3/B3, A4/B4),
the solid lines give the cosine (A) term, and the dashed lines
the sine (B) one.
\label{fig:comp}}
\end{figure*}

\clearpage 


\end{document}